# Engineering Hybrid Physics-Informed Neural Networks for Next-Generation Electricity Systems: A State-of-the-Art Review

Joseph Nyangon[1*]

## Abstract

The integration of machine learning with domain-specific physics is transforming the design, monitoring, and control of modern electricity systems. Physics-informed machine learning (PIML) addresses persistent challenges of data scarcity, limited interpretability, and enforcement of physical laws while delivering computationally efficient and accurate solutions for Industry 4.0 applications. This article presents a state-of-the-art review of hybrid physics-informed neural network (PINN) architectures that fuse data-driven learning with first-principles modelling for real-time diagnostics, digital twins, fault detection, and control optimization across electricity system components. Representative architectures examined include Deep Operator Networks (DeepONets), Fourier Neural Operators (FNOs), Extreme Learning Machine (ELM)–enhanced PINNs, Graph-Based PINNs (PIGNNs), and domain decomposition PINNs, synthesized through comparative analysis of published case studies and simulations. The reviewed evidence demonstrates that PIML delivers robust, scalable, and physically consistent predictions while mitigating parameter sensitivity and dynamic-behavior limitations of purely data-driven approaches. Persistent challenges around computational complexity, uncertainty quantification, and industrial-scale deployment are critically discussed. Collectively, these advances signal a paradigm shift from opaque, black-box data-driven methods toward transparent, white-box, physics-informed strategies for next-generation electricity systems.

## 1. Introduction

### 1.1 Fundamentals of Machine Learning in Electrical Machines and Drives

Machine learning (ML) is increasingly important in the design, control, and optimization of electrical machines and drives. ML techniques offer insights into complex processes without requiring extensive domain knowledge or computational time (Mayr et al., 2017). They are being applied to various aspects, including fault diagnostics (Kudelina et al., 2021; Murphey et al., 2006), performance prediction (Raia et al., 2023), and control optimization (Tom & Febin Daya, 2024; Zhang et al., 2023). ML algorithms can be used for predictive maintenance, quality management, and process control in electric drive production (Mayr et al., 2017). They also enable fast estimation of device performance during design space exploration. In power electronics, ML techniques are crucial for enhancing control and optimization strategies (Bahrami & Khashroum,

[1] Energy Exemplar, Salt Lake City, UT, 84111, USA
* Corresponding author, email: jnyangon@udel.edu

2023). As the industry moves towards Industry 4.0 standards, the integration of ML in electrical machines and drives is expected to become more prevalent, offering improved performance and efficiency across various applications (Kudelina et al., 2021).

As the frontier of power electronics continue to advance with the emergence of ML models, particularly artificial neural networks (ANNs), the fields of complex system identification, control, and estimation in power electronics and motor have gained prominence in recent decades (Bose, 2001, 2005). ANNs have found applications in various aspects of power electronics, including static feedback signal estimation, space vector PWM, and flux vector estimation (Bose, 2020). The integration of AI techniques like fuzzy logic and neural networks has brought new challenges and opportunities to power electronic engineers (Bose, 1994, 2002). However, it has also been established that Convolutional Neural Networks (CNNs) "are one of the many machine learning approaches that have shown great promise in power electronics" for mitigating these changes (Khashroum et al., 2023). Recent developments have focused on employing CNNs for high accuracy in diagnosing bearing faults and detecting winding defects in electric motors (Husebø et al., 2019), classification in manufactured parts (Warren et al., 2021), and pattern recognition in power electronics systems (Abu-Rub et al., 2021; Hao et al., 2023; Brosch et al., 2021). The integration of CNNs with batch normalization has improved fault detection in induction motors (Kumar & Shankar Hati, 2021). Despite challenges such as the need for high-quality data and real-time implementation (Khashroum et al., 2023), CNNs offer significant potential for enhancing reliability and efficiency in electrical machines and drives through automated and accurate fault diagnosis.

Before the deep learning era, ML applications in electric machine drives primarily relied on hardware platforms optimized for high-speed, parallel processing (Mojlish et al., 2017). However, many implementations were constrained using slower, sequentially operated digital signal processors (DSPs). In some cases, multiple DSPs were employed to enhance processing speed. Embedded platforms like field-programmable gate arrays (FPGAs), graphics processing units (GPUs), and chip multiprocessors used for real-time simulations and control of electric machines, were not yet fully developed and thus saw limited use. These platforms offered advantages in simulation acceleration and computational efficiency for implementing ML algorithms (Osta et al., 2019). Even today, hardware limitations remain a significant barrier to ML applications in this field, particularly in industrial settings. Traditional ML techniques, such as support vector machines, neural networks, and random forests, were applied to electric drive production for tasks like predictive maintenance and quality management (Mayr et al., 2017). With the advent of deep learning, there has been a shift towards more advanced algorithms and specialized hardware platforms, enabling improved performance in control and monitoring of electric machine drives (S. Zhang et al., 2023). This transition has led to increased focus on hardware-aware ML modelling and optimization for various applications (Marculescu et al., 2018; Reagen et al., 2017). This hardware bottleneck has hindered the deployment of ML algorithms, leading to suboptimal performance in tasks such as identification and control.

Despite these challenges, recent advancements in ML and FPGAs are driving widespread adoption across various industries. FPGAs offer significant advantages in throughput, latency, and energy efficiency for implementing ML tasks compared to CPUs and GPUs (Kathail, 2020). They excel in applications such as autonomous driving, healthcare, and electronic design automation (Hamolia & Melnyk, 2021; Zhao, 2023). FPGAs provide high parallelism, low latency, and hardware customization abilities, making them ideal for accelerating ML algorithms (Itagi et al., 2021). Researchers are developing optimized FPGA architectures with posit multipliers to enhance performance for ML applications (Elsaid et al., 2022). The integration of ML techniques in FPGA-based systems is also improving control and monitoring of electric machine drives (Zhang et al., 2023). Despite challenges in design considerations and resource utilization, ongoing research aims to address these issues and further enhance FPGA capabilities for ML tasks (Taj & Farooq, 2023). Additionally, ML techniques are increasingly applied across various electric machine drive domains, including control and monitoring of electric machine drives, optimization of power electronics control (Bahrami & Khashroum, 2023), and addressing challenges in power-electronics-dominated grids (Nyangon, 2025b; Abu-Rub et al., 2021). Emerging applications, such as data-driven thermal modelling of power electronics modules (Li et al., 2023) and anomaly detection in power electronics systems (Hossein Rahimighazvini et al., 2024), demonstrate the growing relevance of ML. Looking ahead, deep learning and reinforcement learning algorithms show particular promise, positioning AI-driven techniques as standard tools for the future of electric machine drives (Zhang et al., 2023).

As ML models and embedded systems continue to advance, data-driven approaches are becoming increasingly popular for the high-performance control of electric machine drives. Pruning and quantization are two key techniques that can significantly reduce model size and improve inference speed without substantial accuracy loss (Green et al., 2018; Loureiro et al., 2023). These methods eliminate parameters with minimal impact on a neural network's output and reduce numerical precision, leading to smaller model sizes and faster computations with negligible accuracy loss (Hawks et al., 2021). For instance, quantization-aware pruning has demonstrated improved results compared to using pruning or quantization alone, achieving model size reductions of up to 94% and improving inference times by 73% (Torres-Tello & Ko, 2022). Techniques like deep compression, which integrate pruning, quantization, and Huffman coding, have further reduced model sizes by 35x to 49x without sacrificing performance (Han et al., 2015). These optimizations are particularly valuable for deploying complex neural networks on resource-constrained mobile and embedded devices (Gupta et al., 2022; Li et al., 2023). However, optimizing models for edge applications requires careful balancing of trade-offs between inference time, model size, and predictive accuracy (Loureiro et al., 2023). Scientific ML, parallelization, and high-performance computing further accelerate these processes, particularly for large-scale networks. Physics-informed learning methods can enhance these applications by enabling online estimation of unknown or time-varying system parameters and using neural networks to solve system states rapidly for studying transient dynamics and stability.

### 1.2 Physics-informed Machine Learning (PIML)

PIML integrates physical laws and domain knowledge into ML models, enhancing their accuracy, efficiency, and optimizability (Nghiem et al., 2023; Pateras et al., 2023). PIML offers benefits such as improved data efficiency, physical consistency, and interpretability (Meng et al., 2022). It has been successfully applied in various fields, including weather and climate modelling (Nyangon, 2024), tribology (Marian & Tremmel, 2023), and fluid mechanics (Sharma et al., 2023). PIML approaches can be broadly categorized based on how physics information is derived and incorporated into the learning process (Pateras et al., 2023). Common implementations include physics-informed neural networks and physics-informed graph networks (Shukla et al., 2022). While PIML shows promise in addressing complex multiscale problems, challenges remain in developing robust and reliable models for large-scale engineering applications (Asef & Vagg, 2024).

### 1.2.1. Physics-Informed Neural Networks (PINNs)

PINNs are a deep learning framework that integrates physical laws described by parametrised partial differential equations (PDEs) into neural network architectures (Nyangon, 2025a; Raissi et al., 2019). This is achieved by incorporating the residuals of underlying differential equations as additional loss terms during training, enabling these networks to either encode physical laws or infer them from data. A common formulation of PINNs combines a data loss with a physics loss (Figure 1). One simple expression is:

$$\mathcal{L}(\theta) = \frac{1}{N}\sum_{i=1}^{N} \mid u(x_i) - u_\theta(x_i) \mid^2 + \lambda \, \frac{1}{M}\sum_{j=1}^{M} \mid \mathcal{N}\left[u_\theta\left(x_j\right)\right] \mid^2$$

Here $u(x_i)$ are the observed (or measured) data points, $u_\theta(x_i)$ is the prediction from the neural network parameterized by $\theta$, $\mathcal{N}[u]$ represents a differential operator enforcing the underlying physical law (e.g., a PDE), $N$ and $M$ are the numbers of data and collocation points, respectively, and $\lambda$ is a hyperparameter balancing the data fidelity with the physics constraint.

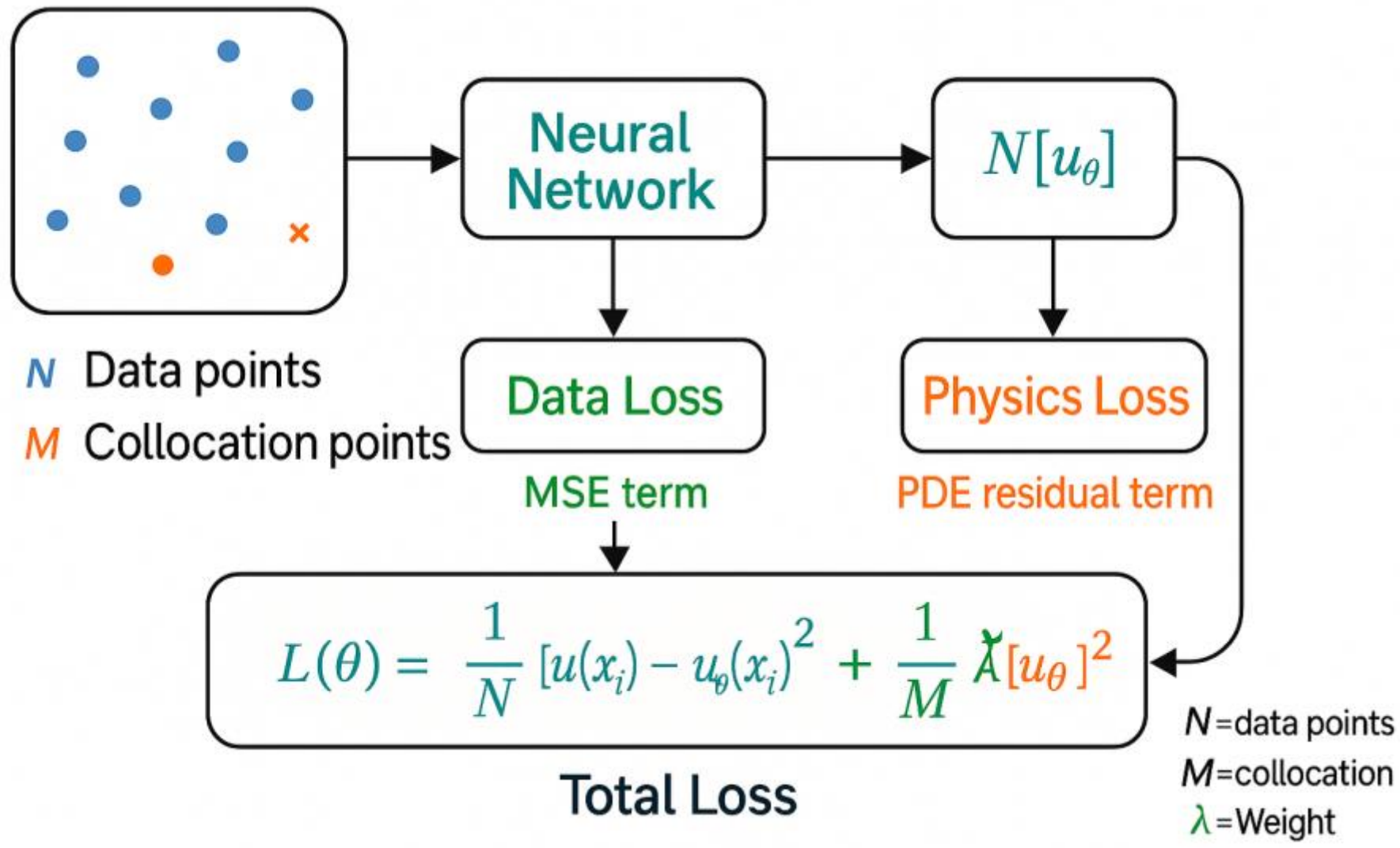

Figure 1: PINNs loss formation

This loss function not only encourages the model to fit the data but also ensures it adheres to the underlying physical laws. While PINNs have shown promise as function approximators, their application to problems exhibiting highly nonlinear, chaotic, or multiscale behaviours has exposed significant challenges. These include issues with stability, convergence, and gradient pathologies (Wang et al., 2021). Recent studies have tackled these challenges by developing specialised activation functions, advanced optimization techniques, and more refined loss function structures (Antonion et al., 2024; Cuomo et al., 2022). Such approaches have enabled PINNs to model complex systems, including the chaotic dynamics of the Lorenz system, the intricate behavior of the Kuramoto–Sivashinsky equation in its chaotic domain, and the turbulent regimes governed by the Navier–Stokes equations. Nevertheless, these targeted solutions often lack broad applicability, as they tend to address isolated challenges rather than ensuring robust performance across diverse scenarios. Consequently, improving the generalizability and convergence properties of PINNs remains a vibrant area of ongoing research (Raissi et al., 2019; Shukla et al., 2022).

The Navier–Stokes equations, for example, form a cornerstone of fluid dynamics by describing the motion of fluid substances. For an incompressible, Newtonian fluid, they are typically expressed as follows:

Continuity equation (mass conservation):

$\nabla \cdot u = 0$

Momentum equation (conservation of momentum):

$$\rho \frac{\partial \mathrm{u}}{\partial \mathrm{t}} + (\mathrm{u}.\nabla)u = -\nabla p + \mu \nabla^2 u + f$$

Here, ρ denotes the fluid density, u the velocity vector field, p the pressure, μ the dynamic viscosity, and f represents body forces per unit volume. In index notation, the momentum equation becomes:

$$\frac{\partial u_i}{\partial t} + u_j \frac{\partial u_i}{\partial x_j} = -\frac{1}{\rho}\frac{\partial p}{\partial x_i} + \nu \frac{\partial^2 u_i}{\partial x_j \partial x_j} + f_i,$$

where $u_i$ is the $i$-th component of the velocity vector $\mathbf{u}$, $p$ is the pressure, $\rho$ is the fluid density, $\nu$ is the kinematic viscosity, $f_i$ represents the $i$-th component of any external body force, and $x_j$ denotes the $j$-th spatial coordinate. For compressible flows, the equations are modified to account for density variations. The continuity equation is then given by:

$$\frac{\partial \rho}{\partial t} + \nabla \cdot (\rho \mathbf{u}) = 0.$$

and the momentum equation is:

$$\frac{\partial (\rho \mathbf{u})}{\partial t} + \nabla \cdot (\rho \mathbf{u} \otimes \mathbf{u}) = -\nabla p + \nabla \cdot \boldsymbol{\tau} + \rho \mathbf{f},$$

where u is the velocity vector, p is the pressure, ρ is the density, f represents external body forces, and τ is the viscous stress tensor. The viscous stress tensor τ for a Newtonian fluid defined as:

$$\tau_{ij} = \mu \left( \frac{\partial u_i}{\partial x_j} + \frac{\partial u_j}{\partial x_i} \right) - \frac{2}{3} \mu \, \delta_{ij} \, \nabla \cdot \mathbf{u},$$

Here, $\tau_{ij}$ is the $(i,j)$-th component of the stress tensor, $\mu$ is the dynamic viscosity, $\delta_{ij}$ is the Kronecker delta (which is 1 if $i = j$ and 0 otherwise), $\nabla \cdot \mathbf{u}$ is the divergence of the velocity field.

These formulations encapsulate the essential physics of fluid dynamics and provide a rigorous foundation for analyzing a wide range of flow problems, including in electric machines and drives.

### 1.2.2. Physics-Informed Graph Neural Networks (PIGNNs)

PIGNNs have emerged as a robust framework for addressing complex dynamical systems by combining the intrinsic advantages of graph neural networks with physics-based constraints. In this approach, the network architecture integrates prior knowledge of physical laws to enforce consistency in predictions, ensuring that outputs remain in alignment with fundamental governing principles (Zhang et al., 2025; Thangamuthu et al., 2023). PIGNNs are particularly effective in handling irregular meshes, accommodating longer time steps, and adapting to varying initial and boundary conditions. By leveraging these attributes, the model achieves improved accuracy and enhanced data efficiency when simulating system dynamics, including conserved quantities (Liu & Pyrcz, 2023). Furthermore, recent innovations—such as the incorporation of causality in model training (Zhang et al., 2023) and the integration of Bayesian inference for uncertainty quantification (Linka et al., 2022)—have expanded the applicability of PIGNNs across multiple scientific domains, including structural dynamics, control problems, and nonlinear system analysis (Antonelo et al., 2024). This method not only improves performance in forward and inverse problem-solving but also offers a promising pathway for zero-shot optimizability to larger system sizes.

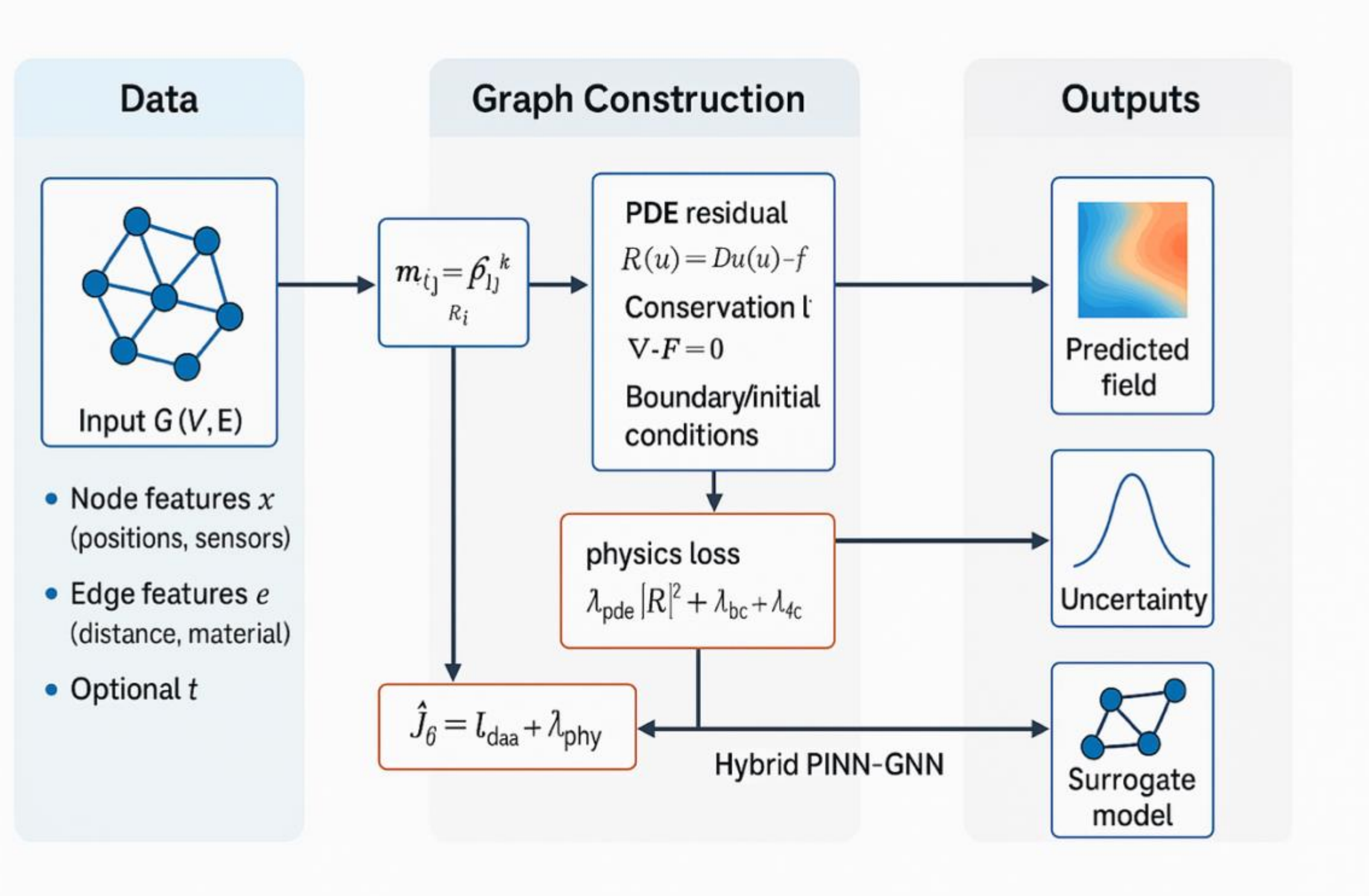

Figure 2: Graphical illustration of PIGNNs interdependencies (author's illustration)

*a) Graph Representation and Data Integration*

The PIGNN framework leverages graph structures to model complex interdependencies inherent in physical systems (Figure 2). Graphs, comprising nodes and edges, effectively represent spatial relationships and connectivity, enabling the network to aggregate local information through various convolutional and attention-based mechanisms (Sen et al., 2024). Within this structure, the adjacency matrix—either informed by domain expertise or self-learned—plays a critical role in quantifying the relationship among system components. Notably, the method is designed to address data limitations, such as missing or incomplete input data. In scenarios where key variables, such as inlet vessel information in fluid dynamics, are absent, model accuracy significantly deteriorates due to the loss of influential data (Wang et al., 2021). This challenge underscores the necessity of precise data integration and highlights areas for future research. By embedding physics constraints directly into the network's training process, PIGNNs mitigate the risk of extrapolating beyond the learned data range and enhance the model's interpretability and predictive capability. Overall, this integrated approach presents a promising advancement in subsurface production forecasting and other applications, offering an effective balance between computational efficiency and physical fidelity (Cuomo et al., 2022; Shukla et al., 2022).

### 1.2 Scope and Contribution

The motivation for applying PIML in the design and control of electrical machines and drives is driven by the need to reconcile data-driven approaches with the immutable laws of physics. This paper provides a comprehensive state-of-the-art review of recent advances in applying PIML to the design, control, and monitoring of electrical machines and drives. By integrating physical laws with ML models, as demonstrated by Meng et al. (2022) and Xu et al. (2023), the approach addresses challenges related to data scarcity and model generalizability while ensuring physical plausibility. Furthermore, advanced methods such as domain decomposition and neural operator learning (Pateras et al., 2023) underscore PIML's capability to tackle high-dimensional problems that traditional numerical methods often find intractable. The incorporation of these methods enhances the rapid optimization of complex geometries and control strategies, thereby improving performance metrics such as torque, efficiency, and overall system reliability in electrical machine applications.

This paper presents a state-of-the-art review and categorization of recent advances in applying PIML methods to the design, control, and monitoring of electrical machines and drives. To this end, the review is organized around three main themes: the underlying motivation for adopting PIML, the integration of physical knowledge into ML frameworks, and the diverse methodological approaches employed. In addition, current challenges—such as data scarcity, optimizability, and the computational expense associated with deep learning models (Hao et al., 2023)—are discussed alongside potential research opportunities for real-time deployment in embedded systems. The review also highlights applications ranging from system identification and control optimization to the development of digital twins (Nghiem et al., 2023; Xu et al., 2023), thereby offering a robust perspective on how PIML can enhance the design and control of modern electrical machines and drives. This review aims to provide both a tutorial framework for new researchers and a detailed reference for practitioners seeking to deploy PIML in high-dimensional, real-time applications within electrical drives.

### 1.3 Outline of the Study

This paper is structured as follows. Section 1 introduces the fundamentals of PIML applications and methods for electrical machines and drives, alongside a review of key industrial machines and their associated challenges. Sections 2 and 3 presents state-of-the-art review of PIML techniques applied to these systems. Section 4 discusses the challenges and future trends for electrical machines and drives enabled by advanced ML algorithms, and Section 5 concludes the paper. This review assumes familiarity with basic ML concepts, given its widespread adoption and the extensive literature available. Consequently, it serves as a practical guide to the successful and diverse applications of ML models within the realm of electrical machines and drives.

## 2. Fundamentals of PIML for Electrical Machines and Drives

### 2.1 Overview of Electrical Machines and Drives

Electrical machines and drives are essential components in energy conversion systems, transforming electrical energy into mechanical energy and vice versa (De Doncker et al., 2011). At their core, these systems encompass a range of devices—from AC, DC, synchronous, to induction machines—each tailored for specific applications through unique operational characteristics and performance benefits (Aissaoui, 2018). Fundamental terminologies such as power, torque, speed, efficiency, and control, underpin the understanding of these devices, where concepts like Faraday's law and the Lorentz force elucidate the conversion processes between electrical and mechanical energy. In motors, electromagnetic interactions produce rotational motion, whereas in generators the same principles facilitate the conversion of mechanical motion into electrical output. Modern drive technologies, including variable frequency drives, vector control, and direct torque control, further refine the performance and efficiency of these systems by enabling precise regulation of speed, torque, and direction (Chen & Tong, 2023). This intricate interplay between classification, operating principles, and control strategies underlines the technological advancements that continue to drive improvements in performance and application versatility.

Design considerations for electrical machines and drives focus on key parameters such as efficiency, power density, and dynamic response across four areas (Figure 3): (a) industrial and general-purpose applications, (b) high-performance and precision control applications, (c) energy-efficient and modern motor technologies, and (d) specialized and harsh environment applications, aligning machine capabilities with specific application demands (Boldea et al., 2018). These design imperatives are evident in a wide range of applications—from electric vehicles and industrial automation to renewable energy systems—where integration with modern control systems and smart grid technologies is increasingly prevalent (Nyangon & Darekar, 2024). For instance, industrial settings often favor robust induction motors for their durability, while permanent magnet motors are preferred in electric vehicles for their improved efficiency and power density. Simultaneously, emerging trends highlight advancements in materials, digital control, and energy-efficient technologies, suggesting a future where even higher performance and adaptability become standard. Innovations in design and the evolving role of drive systems are poised to further enhance operational reliability and sustainability, ensuring that electrical machines continue to meet the growing demands of modern energy conversion challenges (Agamloh et al., 2020; Biasion et al., 2021).

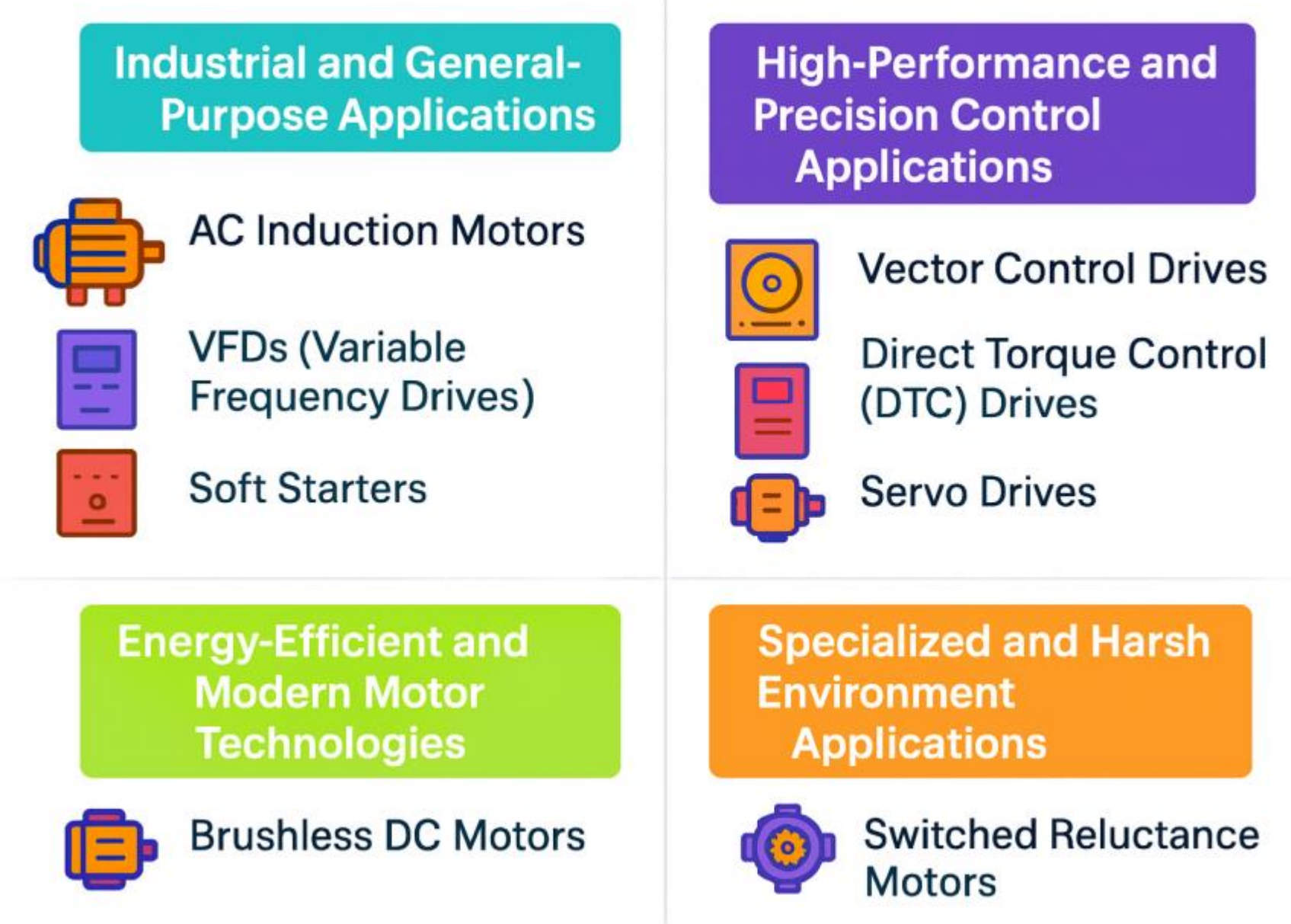

Figure 3: Categorization of electrical machines and drives by application and performance characteristics

### 2.1.1. Industrial and General-Purpose Applications

AC induction motors are widely employed across industrial and general purpose applications due to their simple construction, inherent robustness, and favorable cost–performance trade-offs across domestic, automotive, petrochemical, and oil-and-gas settings (Chen & Tong, 2023; Faiz et al., 2017). Three-phase stator windings coupled to squirrel-cage rotors enable efficient energy conversion under wide load variability, while closed-loop control ensures accurate torque, speed, and position regulation with reversible motor–generator operation—critical for resilience and lifecycle economics relative to brushed alternatives (Bachchhav et al., 2017; Boldea et al., 2018). Recent. Field-weakening extends speed beyond base values and supports scalability from fractional-kilowatt appliances to multi-megawatt compressors, whereas variable-frequency operation suppresses inrush, stabilizes thermal profiles, and improves load-matching efficiency in continuous and cyclic duty. Reliability in hazardous atmospheres mandates explosion-proof or pressurized enclosures to mitigate ignition risks in Class I/Division 1 or Zone 1 applications; compliance engineering adds cost and weight but preserves availability in mission-critical service (Mistry et al., 2024). Ongoing research targets higher power density, reduced losses, and advanced observers for parameter drift, thereby strengthening condition-based maintenance and extending overhaul intervals without eroding capital efficiency (R. Chen & Tong, 2023). Collectively, these attributes—ruggedness, scalability, and controllability—explain the enduring centrality of induction machines in electrified systems spanning pumps, fans, conveyors, and subsea lift where mean-time-between-failure and total cost of ownership dominate specification.

Variable-frequency drives (VFDs) and soft starters provide complementary actuation layers that tune electrical and mechanical stress to duty profiles, improving efficiency, reliability, and asset life at system level. VFDs modulate voltage and frequency to align electromagnetic torque with instantaneous demand, enabling soft-start, load-shedding, and speed optimization in upstream oil and gas (beam pumping, FPSOs), petrochemicals, hydrogen process auxiliaries, HVAC, and transport—where energy savings and harmonics compliance must be balanced against over-torque and partial-load losses (Nyangon & Darekar, 2024; Antunes et al., 2023; Tan et al., 2020). Recent work refines control algorithms and protection logics to harden drives against grid disturbances and non-ideal waveforms, lifting availability in remote or islanded power systems (X. Guo et al., 2024; Kapp et al., 2024). Soft starters, by contrast, regulate input voltage at fixed frequency using thyristor stages to curtail starting current, mitigate torque ripple, and improve input power factor with minimal component count and attractive first cost (Nannen et al., 2022; Deraz & Azazi, 2017). They have scaled across fans, pumps, compressors, mixers, and electric submersible pumps in aerospace, manufacturing, and oil and gas, while research advances—three-phase PWM AC choppers—reduce sensors and harmonic distortion to limit rotor heating and bearing stress (Biot-Monterde et al., 2021; Corral-Hernandez & Antonino-Daviu, 2018; Öksüztepe et al., 2022; Rabbi et al., 2020; Zargari et al., 2017). In combination, VFDs deliver maximal energy optimization and flexibility; soft starters supply cost-effective, reliable starts—together furnishing robust, scalable drive solutions across duty classes.

#### 2.1.2. High-Performance and Precision Control Applications

Permanent-magnet synchronous machines (PMSM) and their associated vector-controlled drives underpin high-performance motion systems in electric vehicles, industrial automation, CNC tooling, and robotics, where strict torque and speed accuracy, low latency, and compact integration are paramount (M. H. Ali et al., 2024; Tripathi & Vaish, 2019). A laminated stator carrying three-phase windings establishes a rotating field that locks synchronously with rotor-mounted permanent magnets, yielding high power density, excellent efficiency, a wide constant-torque range, and inherently smooth torque production essential for contouring and pick-and-place tasks (Tom & Febin Daya, 2024). In this regime, field-oriented control decouples flux and torque channels, enabling linearized current-to-torque behavior and fast dynamic response under load disturbances, while modern estimators preserve precision at partial load and low speed (Zhu et al., 2021). The absence of brushes reduces maintenance and acoustic noise, and the compact electromagnetic package promotes installation in space-constrained end-effectors and aerospace actuators (Tom & Febin Daya, 2024). Recent work integrates machine-learning-assisted parameter identification and on-line thermal models to stabilize bandwidth despite magnet temperature drift and inverter non-idealities, improving transient and steady-state error budgets (Tom & Febin Daya, 2024; Aljafari et al., 2022). These attributes translate directly into higher throughput and tighter tolerance in precision assembly cells, wafer handling, and coordinated multi-axis gantries. Consequently, PMSM platforms, paired with vector control, form the benchmark for energy-efficient servo axes, delivering repeatable micrometer-class positioning and millisecond-scale torque steps while containing switching losses through optimal modulation and advanced observers (Zhu et al., 2021; Aljafari et al., 2022).

Direct-torque-controlled drives complement vector-controlled PMSM systems where near-instantaneous torque steps, minimal parameter dependence, and robust disturbance rejection are required, as in high-speed pick-and-place, robotic joints, and spindle drives for CNC machining (Meesala & Thippiripati, 2020; Foo & Zhang, 2016). By regulating electromagnetic torque and stator flux directly, DTC avoids coordinate transformations and current-loop cascades, achieving superior transient behavior at the expense of characteristic torque ripple and variable switching frequency; contemporary schemes moderate these effects using constant-frequency modulators, refined look-up tables, and virtual voltage space vectors (Hakami et al., 2019). Dual-inverter and multi-level variants suppress acoustic noise and improve efficiency under wide speed ranges and fluctuating loads, consolidating DTC's role in electric vehicles, high-power induction drives, and industrial actuators (Suresh & Rajeevan, 2020; Foo & Zhang, 2016). On the application layer, servo drives integrate these control primitives—field-oriented, DTC, and model predictive control—behind unified motion controllers providing precise speed, acceleration, and position profiles with jerk constraints, feed-forward shaping, and synchronized multi-axis interpolation (Kitayoshi et al., 2023; Flieh et al., 2017). Coupled with industrial networks and IoT gateways, modern servo platforms stream telemetry for condition monitoring, on-line identification, and predictive maintenance, thereby sustaining bandwidth and positioning accuracy as mechanisms wear and thermal states drift (Kitayoshi et al., 2023). Emerging direct-drive architectures and multi-degree-of-freedom stages extend stiffness and back-drivability benefits to collaborative robots and precision metrology whilst reducing backlash, gear noise, and maintenance (Ruderman et al., 2020). Together, these advances deliver repeatable sub-arcsecond positioning, low settling times, and energy-aware motion, enabling higher utilization, safer human-robot collaboration, and consistent quality across automation cells.

#### 2.1.3. Energy-Efficient and Modern Motor Technologies

Synchronous Reluctance Motors (SynRMs) are AC machines without rotor windings or magnets, yielding compact, low-loss designs with high torque density and lower cost (Bianchi et al., 2021; Mbula & Chowdhury, 2017). Recent optimization of rotor saliency and flux-barrier geometry mitigates torque ripple and improves power factor, closing the gap between theoretical and realized performance (Murataliyev et al., 2022). In industrial duty—fans, pumps, mills—SynRMs increasingly outperform induction motors on efficiency, enabling smaller frames, reduced cooling demand, and lower lifetime energy spend (Ozcelik et al., 2019; Donaghy-Spargo, 2016). Their magnet-free bill of materials reduces rare-earth risk and aligns with green-tech procurement goals. When paired with modern vector drives, SynRMs deliver accurate speed control and strong low-speed torque, suiting EV auxiliaries, HVAC, and process control. Ongoing materials and control advances position SynRMs as an emergent topology across energy-intensive sectors (Bianchi et al., 2021; Ozcelik et al., 2019).

Brushless DC (BLDC) motors are permanent-magnet machines that eliminate brushes and commutators, reducing copper loss, friction and maintenance while improving efficiency and power density. Contemporary drive strategies—field-oriented control, direct-torque control and model-predictive control—suppress torque ripple, enhance low-speed smoothness, and minimize switching loss to optimize energy use (Mohanraj et al., 2022; Prabhu et al., 2023). The result is compact actuators with high torque-to-weight ratios for EV traction and auxiliaries, UAVs, robotics, haptics, pumps, and HVAC compressors, where partial-load efficiency is critical. Advances in semiconductor packaging, current sensing and calibration further improve dynamic response and EMI robustness, widening suitability for HVAC, manufacturing and green-tech platforms. Remaining challenges—thermal bottlenecks in dense stators and fault diagnostics in sensor-limited systems—are active areas of research, with promising results in estimation-based protection and enhanced cooling (Shokri & Naderi, 2017). As costs fall and controls mature, BLDC drives will continue to displace induction solutions in efficiency-critical applications (Mohanraj et al., 2022).

### 2.1.4. Specialized and Harsh Environment Applications

Switched reluctance motors (SRMs) exploit variable reluctance and a salient-pole rotor, yielding a simple, magnet-free topology tolerant to heat, vibration and contamination. Their robust laminations, mechanical clearances and thermally resilient windings support sustained operation under high-temperature duty cycles and shock-prone settings typical of EV powertrains, aerospace actuators, wind generator pitch drives and heavy-duty appliances (Li et al., 2019; Xu et al., 2022). Absence of rare-earth magnets eliminates demagnetization risk at elevated temperatures and reduces cost volatility, while improving thermal headroom for higher current densities. Modern control—especially model predictive control—and multiphase configurations suppress torque ripple and acoustic noise, enhance fault tolerance, and stabilize nonlinear behavior across wide speed ranges (Seshadri & Lenin, 2020; Valencia et al., 2021). Collectively, these attributes position SRMs as reliable, efficient alternatives to induction and permanent-magnet machines for harsh-environment duty, with manufacturability and lifecycle economics (Diao et al., 2022). Table 1 summarizes major electrical machines, drives, applications and advantages.

Table 1: Electrical machines, drives, applications and PIML models

| Component | Applications | Advantages | Representatives |
|---|---|---|---|
| AC induction motors | Industrial machinery, HVAC systems, pumps, fans | Robust, cost-effective, high reliability | PINNs for system identification |
| PMSM | Robotics, electric vehicles, precision instrumentation | High efficiency, improved performance, precise control | Hybrid models combining physics-based simulation with deep learning |
| SynRMs | Energy-efficient industrial applications, fans, pumps | Improved efficiency, reduced maintenance costs | PINN-enhanced optimization models |
| SRMs | Harsh environment applications, automotive, industrial systems | Simple design, robust under extreme conditions | Data-driven approaches with embedded physical constraints (PINNs) |
| BLDC | Household appliances, robotics, industrial automation, medical devices | High reliability, precise control, energy-efficient | Physics-informed hybrid deep learning models |
| VFD | AC motor control in industrial, HVAC, and pump applications | Energy efficiency, variable speed control, and reduced power consumption | PINN-based predictive maintenance and control models |
| Vector control drives | Robotics, precision manufacturing, high-performance industrial systems | Enhanced torque and speed regulation, improved dynamic response | Hybrid PINN models for advanced control strategies |

| DTC drives | High dynamic industrial applications, traction control systems | Rapid torque response, improved dynamic performance | Real-time PIML integration for torque and efficiency optimization |
|---|---|---|---|
| Servo drives | Robotics, CNC machinery, automation systems | High precision, fast dynamic response, accurate motion control | Digital twins with embedded PIML for predictive maintenance and control |
| Soft starters | Motor startup applications in industrial and commercial settings | Controlled acceleration, reduced mechanical stress and inrush currents | PINN-based anomaly detection, predictive control models and early fault detection |

## 2.2 PIML and Operator Learning for Electromagnetic Field Analysis of Electrical Machines

Recent advancements in PIML have significantly enhanced its application across diverse scientific fields, including tribology, fluid mechanics, and chemical engineering (Marian & Tremmel, 2023; Sharma et al., 2023; Wu et al., 2023). By embedding physical laws into ML frameworks, PIML offers improved accuracy, interpretability, and transferability over traditional data-driven methods (Karniadakis et al., 2021). Figure 4 illustrates a graphic linking PIML and operator learning to outputs, showing physics/data losses, operator nets, hybrid surrogate, and active-learning feedback for electromagnetic machine analysis and deployment. Key advances include the evolution of PINNs (section 1.2.1) and PIGNNs (section 1.2.2), which enable the resolution of complex multi-physics problems and the discovery of hidden physics (Shukla et al., 2022). Further progress has been made with: (i) enhanced PINN architectures with methods such as Extreme Learning Machines (ELM), (ii) neural operator techniques for learning mappings between function spaces beyond solving partials pointwise, (iii) novel approaches in domain decomposition and modular frameworks, (iv) adaptive optimization, multi-fidelity strategies, and (v) the integration of functional interpolation techniques with PINN for increased data efficiency and stabilized predictions. These improvements have been demonstrated in seismic wave modelling, multi-physics coupling in chemical engineering, and anomaly detection in smart grids (Zideh et al., 2024; Nyangon & Akintunde, 2024; Wu et al., 2023). Table 2 details recent advancements in PIML along with their technical innovations, benefits, and the areas of application.

Table 2: Recent advances in PIML methods: techniques, benefits, and applications

| Recent advancement | Key features & recent developments | Impact and benefits | Applications |
|---|---|---|---|
| Evolution of PINNs | ▪ Adaptive weighting and multi-scale feature extraction.<br>▪ Enhanced loss functions, automatic differentiation. | ▪ Improved PDE convergence<br>▪ Robustness in diverse applications | ▪ Fluid dynamics<br>▪ Heat transfer<br>▪ Structural mechanics |
| Advances in PIGNNs | ▪ Graph-based spatial interaction modelling.<br>▪ Multi-physics interaction modelling to uncover hidden physics. | ▪ Solves complex multi-physics problems<br>▪ Discovers latent physical laws | ▪ Electrical networks<br>▪ Biological networks<br>▪ Material science |
| Enhanced PINN architectures with ELM | ▪ Integrates ELMs to expedite training with randomly weighted hidden layers<br>▪ Analytic output weighting. | ▪ Reduces training time, lower computation costs<br>▪ Improved generalization. | ▪ Real-time simulations<br>▪ Process control<br>▪ Engineering prototyping |
| Neural operator techniques | ▪ Learns mapping between function spaces<br>▪ DeepONet and Fourier Neural Operators (FNO) that capture global solution behaviors. | ▪ Efficient high-dimensional problem solving<br>▪ Captures global solution behaviors | ▪ Weather prediction<br>▪ Climate modelling<br>▪ Uncertainty quantification |
| Domain decomposition and modular frameworks | ▪ Subdomains solved by modular PINNs.<br>▪ Parallel training for heterogenous systems. | ▪ Scalable for large problems<br>▪ Better boundary or interface handling | ▪ Structural analysis<br>▪ Geophysical simulations<br>▪ Industrial processes |
| Adaptive optimization and multi-fidelity strategies | ▪ Adaptive learning rate and sampling.<br>▪ Multi-fidelity data integration | ▪ Efficient training and accurate models.<br>▪ Cost-effective simulations | ▪ Aerospace design<br>▪ Automotive engineering<br>▪ Environmental modelling |
| Functional interpolation with PINNs | ▪ Couples PINNs with interpolation methods (e.g., kernel-based techniques) to enhance performance on sparse or noisy datasets.<br>▪ Stabilizes predictions with functional priors | ▪ Increased data efficiency and stability of predictions<br>▪ Reliable results in under-sampled or noisy data environments. | ▪ Biomedical imaging<br>▪ Geostatistics<br>▪ Remote sensing |

Figure 4 illustrates a workflow linking PIML and operator learning to predict electromagnetic fields, drive optimization and real-time deployment, featuring active-learning feedback from outputs.

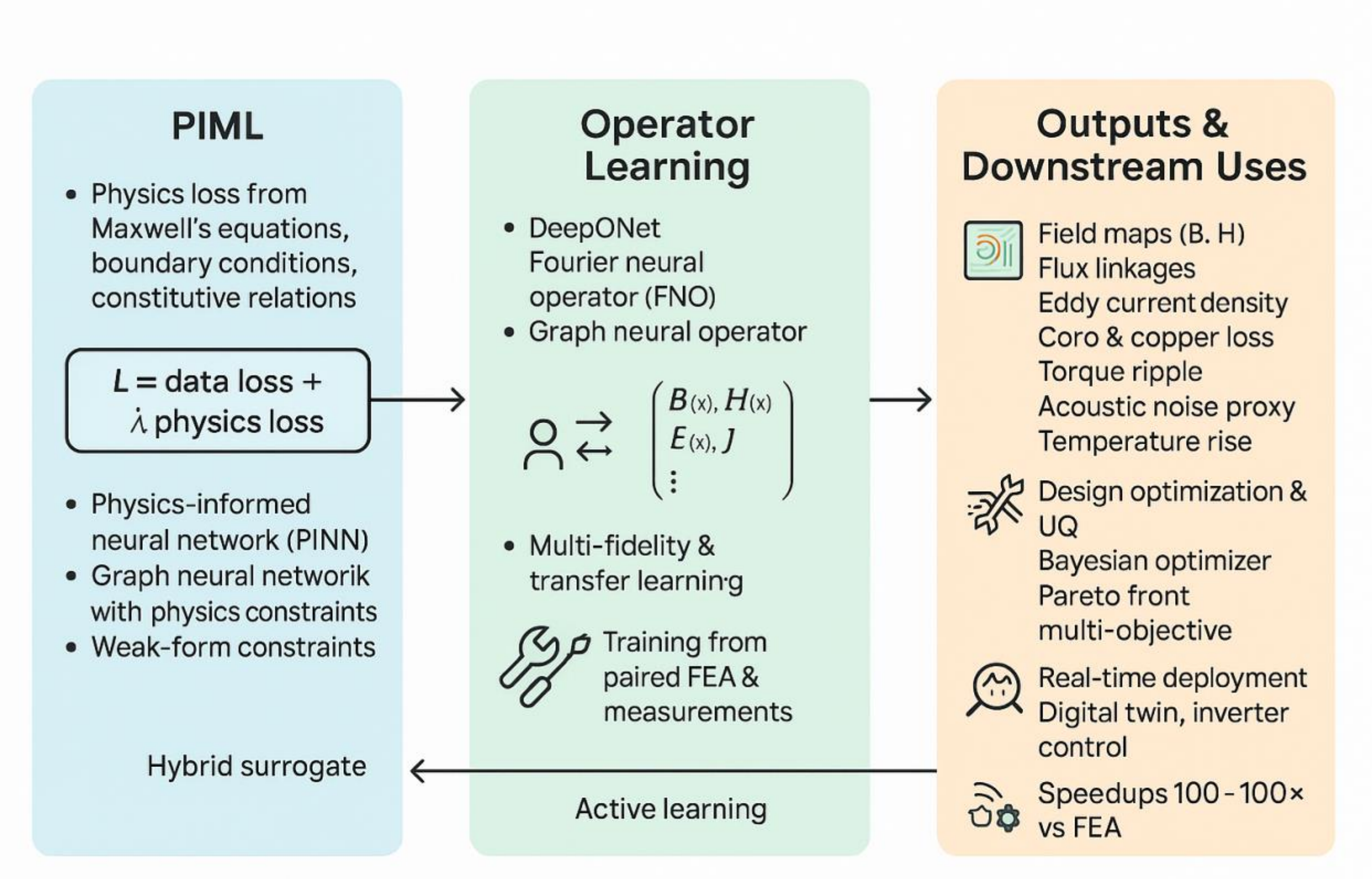

Figure 4: PIML and operator learning for electromagnetic analysis of electrical machines

### 2.2.1. Enhanced PINN Architectures

Recent developments in enhanced PINNs have transformed these models into more robust and efficient tools for tackling complex engineering problems (Nyangon, 2025a). A major evolution has been the hybridization of traditional PINNs with ELMs, which substitute iterative gradient-based updates with a single-step least-squares solution. This hybrid approach, as demonstrated by Joshi et al. (2024) and Khademi & Dufour (2024), significantly reduces training times while maintaining high accuracy, especially in linear or mildly nonlinear systems. Additionally, theory-constrained PINNs have emerged by embedding explicit physical laws, such as first-order shear deformation theory, directly into the loss function (Zhang et al., 2023). This innovation ensures that predictions adhere more closely to established physical behavior, thus narrowing the solution space and enhancing reliability in scenarios with complex boundary conditions. The integration of physics-based regularization within the training loss further mitigates overfitting and improves model fidelity.

Furthermore, advancements in loss function design and optimization have played a crucial role in enhancing PINN performance. Multi-objective loss functions now balance data-driven and physics-based terms through adaptive weighting and dynamic tuning, which alleviates issues like gradient imbalances during training. These improvements not only accelerate convergence but also ensure that the network effectively learns both empirical data and embedded physical constraints. Figure 5 shows enhanced PINNs combining PINN–ELM architecture and physics module with multi-objective loss—Pareto optimization, adaptive weighting, Lagrangian methods—producing solutions, parameter inference, uncertainty, and downstream surrogates. Finally, scalability and robustness have been markedly improved by leveraging domain decomposition techniques, such as those found in Finite Basis PINNs (FBPINNs) (Moseley et al., 2023). A simple FBPINNs formulation augments a neural network's representation with Fourier basis functions. For example, one may represent the solution $u(x)$ as

$$u(x) \approx \sum_{k=1}^{K} [a_k \cos(2\pi\omega_k x) + b_k \sin(2\pi\omega_k x)] + f_\theta(x),$$

where:

- The Fourier series component $\sum_{k=1}^{K}[a_k \cos(2\pi\omega_k x) + b_k \sin(2\pi\omega_k x)]$ captures the global, low-frequency behavior.
- $f_\theta(x)$ is a neural network (with parameters $\theta$) that accounts for local high-frequency corrections or residual features.
- $a_k$, $b_k$ are Fourier coefficients and $\omega_k$ are predefined frequencies.

The training loss then combines data fitting with physics constraints:

$$\mathcal{L}(\theta, a, b) = \frac{1}{N} \sum_{i=1}^{N} | u(x_i) - \sum_{k=1}^{K} [a_k \cos(2\pi\omega_k x_i) + b_k \sin(2\pi\omega_k x_i)] - f_\theta(x_i) |_2 + \lambda \frac{1}{M} \sum_{j=1}^{M} | \mathcal{N} \left[ \sum_{k=1}^{K} [a_k \cos(2\pi\omega_k x_j) + b_k \sin(2\pi\omega_k x_j)] + f_\theta(x_j) \right] |_2.$$

Here, $u(x_i)$ are the observed data, $\mathcal{N}[\cdot]$ denotes a differential operator imposing the underlying physics (e.g., from a PDE), $N$ and $M$ are the numbers of data and collocation points, respectively, and $\lambda$ is a hyperparameter that balances data fidelity and physics enforcement.

This expression illustrates how FBPINNs combine traditional Fourier-based representations with the flexibility of neural networks to learn solutions that adhere to both data and physics constraints. By dividing the problem domain into smaller subdomains and using modular network architectures, these models can efficiently tackle multi-scale and heterogeneous issues. The modular approach allows for parallel training and smooth transitions across subdomain interfaces, enhancing computational efficiency and model robustness. These advancements demonstrate how PINNs have evolved through strategic hybridization, theory-driven constraints, optimized loss formulations, and domain decomposition. Consequently, PINNs have broadened their applicability in various scientific and engineering fields, offering faster and more precise solutions to real-world problems.

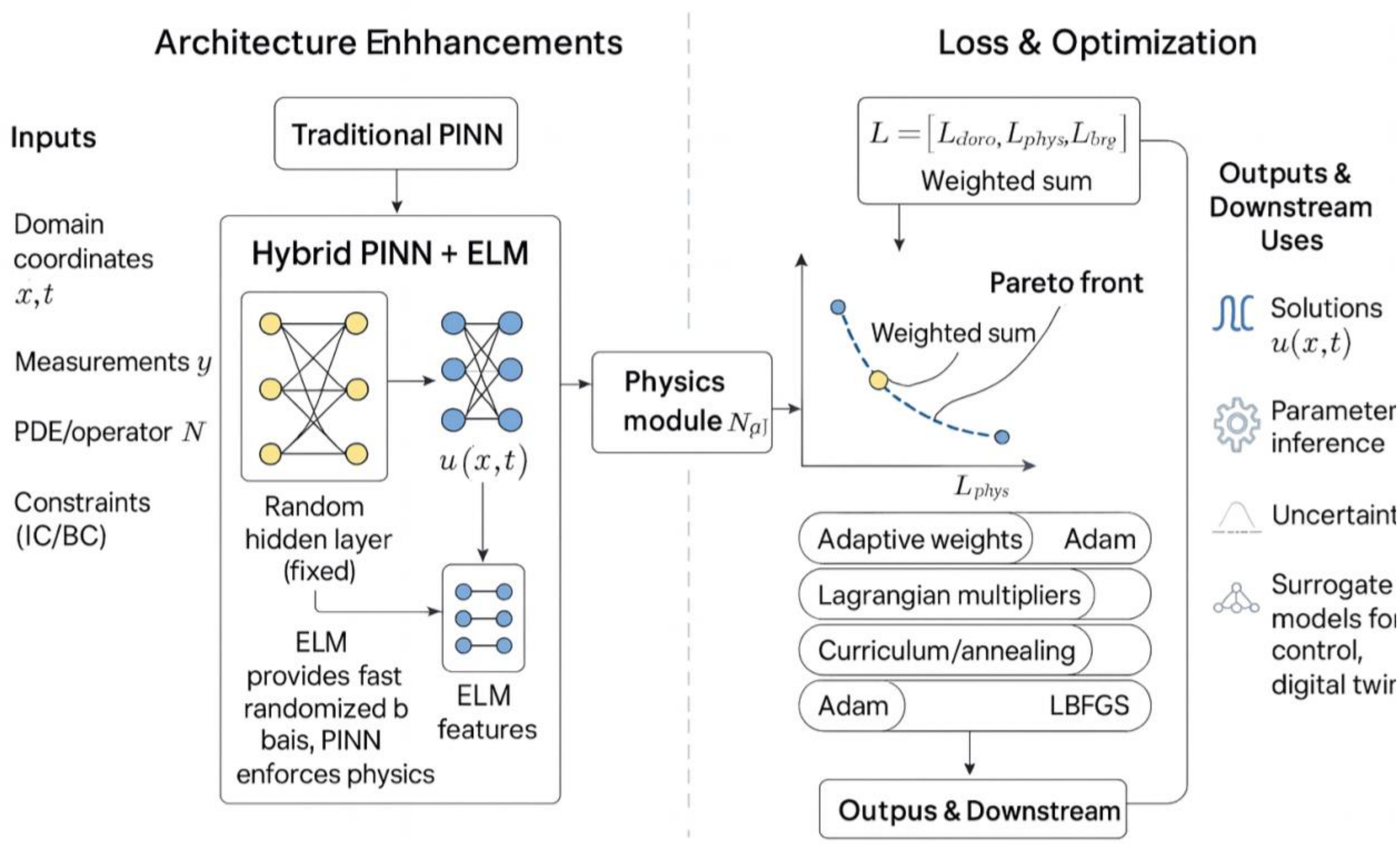

Figure 5: Enhanced PINN architectures with elms and multi-objective loss optimization

### 2.2.2. Domain Decomposition and Modular Approaches

Domain decomposition and modular approaches in PIML have evolved significantly with recent innovations such as disjointed PINNs and extended PINNs (XPINNs), which partition complex, multi-scale problems into smaller subdomains to accelerate training and inference while enhancing scalability in the presence of sharp gradients or heterogeneous materials. In this framework, each subdomain is addressed by an individual neural network with optimally selected hyperparameters, thereby facilitating parallel computation and reducing training costs (Jagtap & Karniadakis, 2020). By enabling arbitrary partitioning in both space and time, XPINNs overcome the limitations of traditional PINNs and even the more conservative PINN (cPINN) approaches, as they allow for flexible representation of localized phenomena and improved handling of discontinuities in material properties (Gu et al., 2024). The modular nature of this method permits the integration of diverse multi-physics data, and the disjointed structure supports robust interface coupling through continuity conditions, ultimately leading to enhanced predictive accuracy and convergence (Wu et al., 2024; Shukla et al., 2022). Consequently, these advances not only streamline the computational process but also provide a scalable solution for solving nonlinear partial differential equations in complex domains, marking a significant step forward in the application of physics-informed neural networks for practical engineering challenges.

### 2.2.3. Operator Learning and Neural Operators

Physics-informed neural operator (PINO) frameworks represent a major leap forward in PIML. PINOs learn mappings between entire function spaces rather than individual pointwise solutions, thereby offering discretization invariance – once trained, these models are not tied to a specific grid or mesh and can be evaluated across various discretizations without retraining (Rosofsky et al., 2023). This flexibility is particularly advantageous in applications such as fluid dynamics and weather forecasting, where simulation grids must adapt to different resolution requirements (Nyangon, 2025b). Moreover, PINOs facilitate rapid surrogate modelling by approximating the entire solution operator, which dramatically reduces the computational time compared to traditional numerical methods. In this context, the Physical Invariant Attention Neural Operator (PIANO) has emerged as a novel variant that integrates physical invariants into the model architecture, thereby enhancing performance by effectively deciphering underlying physical laws (Zhang et al., 2025; Li et al., 2024). By embedding discretization invariance and fast surrogate modelling within their framework, these neural operators provide robust and efficient tools for solving complex, multi-scale physical problems.

Enhanced generalization across parameter spaces and scalability to multi-physics problems further underscore the transformative potential of neural operator frameworks. By capturing the intrinsic mapping between functions, PINOs generalize well to new, unseen scenarios, allowing for interpolation and even extrapolation across varying physical conditions without extensive retraining. This robustness is essential for applications involving high-dimensional parameter spaces, where traditional methods falter. Additionally, the ability to integrate physics-based constraints, such as conservation laws, enables these operators to scale efficiently to complex multi-physics problems, including advanced aerospace simulations and climate modelling. The incorporation of techniques like the fast Fourier transform (FFT) in architectures such as the Fourier Neural Operator (FNO) enhances computational efficiency and prediction accuracy (Rosofsky et al., 2023). Collectively, these innovations demonstrate that both PINOs and PIANO offer promising pathways for accelerating simulation processes while maintaining high fidelity in modelling complex, real-world phenomena.

### 2.2.4 Neural Networks for the Design and Optimization of Electrical Machines

Adaptive optimization and multi-fidelity frameworks significantly enhance PIML by addressing critical challenges in training PINNs. A key improvement is dynamic loss balancing through adaptive weight tuning, which optimally adjusts the contributions of data loss and physics-based residual loss. By employing multi-objective optimization techniques such as gradient normalization, training processes are stabilized, preventing one loss component from dominating, particularly when data is sparse or noisy (Rohrhofer et al., 2023). Transfer learning techniques further improve convergence efficiency by leveraging pre-trained models, allowing PINNs to optimize across different physical systems. This method reduces sample complexity by fine-tuning networks with low-fidelity or sparse datasets based on high-fidelity simulations, thus accelerating training while preserving predictive accuracy (Xu et al., 2023). Moreover, robust optimization algorithms incorporating second-order techniques, adaptive learning rate schedules, and meta-learning approaches mitigate non-convexity issues in loss landscapes, ensuring stable convergence even in high-dimensional problems (Escapil-Inchauspé & Ruz, 2023).

Multi-fidelity data integration plays a crucial role in improving PIML model accuracy and robustness. By seamlessly combining heterogeneous data sources, such as high-fidelity experimental results with lower-fidelity approximations, predictive performance is enhanced while computational costs are minimized (Liu & Wang, 2019). This approach enables the use of inexpensive, low-fidelity data to guide training while sparse high-fidelity measurements correct errors and quantify uncertainties at multiple levels. Additionally, conservative PINNs (cPINNs) and XPINNs are two basic methods that use the domain decomposition framework, further improving scalability in large-scale problems (Shukla et al., 2022). To formulate a domain decomposition approach for cPINNs applied to electric machine systems that obey conservation laws, assume a global domain $\Omega$ is split into $N$ subdomains,$\Omega = \cup_{i=j}^{N} \Omega_i$, and a separate neural network (or a partition of the network) approximates the solution in each subdomain. The overall loss function can then be built by summing contributions from the residuals (i.e., the conservation laws), the enforcement of boundary conditions, and the continuity conditions across interfaces between subdomains.

Assume that the underlying conservation law (for instance, conservation of charge or energy) can be written abstractly as

$$\mathcal{N}[u(\mathrm{x}, t)] = 0, \quad (\mathbf{x}, t) \in \Omega,$$

where $\mathcal{N}$ is a differential operator (which, in an electric machine system, could arise from Maxwell's equations or other energy/charge conservation equations). In each subdomain $\Omega_i$, let

$$u_i(\mathrm{x}, t; \theta_i)$$

be the neural network approximation of the solution. Then the overall loss can be written as

$$\min_{\{\theta_i\}} \mathcal{L}(\{\theta_i\}) = \sum_{i=1}^{N} \left[ \mathcal{L}_{\mathrm{res}}^{(i)} + \lambda_{\mathrm{BC}} \, \mathcal{L}_{\mathrm{BC}}^{(i)} \right] + \lambda_{\mathrm{int}} \sum_{(i,j) \in \mathcal{I}} \mathcal{L}_{\mathrm{int}}^{(i,j)},$$

where:

a) Residual loss in each subdomain (conservation law enforcement):

$$\mathcal{L}_{\mathrm{res}}^{(i)} = \frac{1}{N_i} \sum_{k=1}^{N_i} \| \mathcal{N}[u_i(\mathrm{x}_k, t_k; \theta_i)] \|^2,$$

with $(\mathbf{x}_k, t_k)$ denoting collocation points in subdomain $\Omega_i$.

b) Boundary condition loss in each subdomain:

$$\mathcal{L}_{\mathrm{BC}}^{(i)} = \frac{1}{M_i} \sum_{l=1}^{M_i} \| u_i(\mathrm{x}_l, t_l; \theta_i) - g(\mathrm{x}_l, t_l) \|^2,$$

where $g$ is the prescribed boundary condition function at the boundary points $(\mathbf{x}_l, t_l)$.

c) Interface loss (enforcing continuity across subdomains):

$$\mathcal{L}_{\mathrm{int}}^{(i,j)} = \frac{1}{P_{ij}} \sum_{m=1}^{P_{ij}} \| u_i(\mathrm{x}_m, t_m; \theta_i) - u_j(\mathrm{x}_m, t_m; \theta_j) \|^2,$$

with the sum taken over collocation points $(\mathbf{x}_m, t_m)$ along the interface $\Gamma_{ij} = \partial\Omega_i \cap \partial\Omega_j$ between subdomains $\Omega_i$ and $\Omega_j$.

Here, $\lambda_{\mathrm{BC}}$ and $\lambda_{\mathrm{int}}$ are penalty weights that balance the contributions from the boundary and interface terms.

Despite these advancements, PINNs remain difficult to train due to sensitivity in loss weight selection. However, research demonstrates that system parameterization significantly impacts loss scaling, requiring adaptive strategies for optimal performance (Rohrhofer et al., 2023). Advanced frameworks incorporating domain adaptation techniques and automatic loss weighting schemes hold the potential to streamline training, making PIML a powerful tool for applications in geophysics, materials modelling, and fluid mechanics (Guo et al., 2024).

### 2.2.5. Integration with Functional Interpolation Techniques

PIML integrated with functional interpolation techniques, such as the Theory of Functional Connections (TFC), is advancing computational methods by enforcing boundary and initial conditions analytically. TFC offers closed-form expressions that satisfy constraints exactly, thereby eliminating the need for "soft" imposition via penalty terms (Laghi et al., 2023; Leake & Mortari, 2020). The core idea behind the TFC is to recast a constrained problem into an unconstrained one by "building in" the constraints exactly into the solution form. In TFC, one typically expresses the unknown function as the sum of two parts: (a) a particular "base" function that satisfies the constraints, and (b) a free function (often denoted $g(x)$) multiplied by "blending" functions that vanish at the constraint locations. A general TFC formulation for a problem with linear constraints can be written as:

$$f(x) = g(x) + \sum_{i=1}^{m} \psi_i(x)(c_i - \mathcal{L}_i\{g(x)\}),$$

where:

- $\mathcal{L}_i\{f(x)\} = c_i$ for $i = 1, \ldots, m$ represent the linear constraints (for example, boundary or initial conditions),
- $c_i$ are the prescribed values,
- $g(x)$ is an arbitrary free function, and
- $\psi_i(x)$ are carefully chosen functions that satisfy the conditions

$$\mathcal{L}_j\{\psi_i(x)\} = \delta_{ij}, \quad j, i = 1, \ldots, m,$$

with $\delta_{ij}$ being the Kronecker delta. This guarantees that when the operator $\mathcal{L}_j$ is applied to $f(x)$, the corrective terms exactly "cancel out" any deviation from the constraint caused by the free function $g(x)$. This analytical enforcement restricts the neural network to a subspace of functions that inherently meet the problem's constraints, reducing the search space and minimizing the risk of converging to spurious or non-physical solutions. Furthermore, by removing the competing objectives between data fit and constraint satisfaction, the optimization process experiences enhanced convergence rates and improved numerical stability, particularly in stiff or high-dimensional scenarios (De Florio et al., 2022). Such benefits not only bolster model reliability but also ease the computational burden during training.

Furthermore, integrating TFC with PINNs promotes improved scalability and generalization across complex domains. With constraints handled analytically, models can generalize more effectively over varied discretizations and domain geometries, a vital advantage for multi-scale and multi-physics challenges (Koyama et al., 2024). This approach has demonstrated success in fields such as fluid mechanics, chemical engineering, and tribology, where predictive performance and interpretability are crucial (Marian & Tremmel, 2023). Recent frameworks—like the Extreme Theory of Functional Connections (X-TFC)—enhance these capabilities by combining shallow neural networks with random feature techniques, ensuring robust solutions even with limited or perturbed data (Schiassi et al., 2021). Collectively, these developments signify a shift towards more accurate, interpretable, and computationally efficient models in solving differential equations.

### 2.3 The Benefits of PIML in Engineering Applications

By incorporating domain-specific knowledge directly into the learning process, PIML significantly enhances data efficiency, thereby reducing the reliance on vast datasets while maintaining high fidelity in modelling complex phenomena (Mackay & Nowell, 2023; Wu et al., 2023). This integration not only improves the interpretability of the models but also ensures that predictions remain consistent with known physical principles, which is crucial for applications ranging from fluid mechanics to structural integrity (Marian & Tremmel, 2023; Sharma et al., 2023). Furthermore, the ability of PIML techniques—such as PINNs and PIGNNs—to implicitly capture underlying dynamics presents a promising avenue for discovering hidden physics in systems with sparse or noisy data, thus offering a robust framework for addressing multi-physics challenges.

In addition, PIML provides substantial benefits by streamlining the simulation of complex engineering processes through a reduction in computational overhead and enhanced generalizability. The methodology allows for the seamless integration of multimodality and multi-fidelity data, ensuring that even limited datasets yield reliable and physically consistent predictions (Wu et al., 2023). This capacity to incorporate prior physical knowledge reduces the need for exhaustive data re-learning, thereby expediting the modelling of transport processes, chemical reactions, and structural behavior (Kargah-Ostadi et al., 2024; Shukla et al., 2022). By doing so, PIML not only advances the precision of predictive models but also fosters innovative solutions for design, maintenance, and asset management in engineering, marking a significant leap forward in the field (Zhao et al., 2020).

Table 3 summarizes the benefits of PIML in engineering applications, demonstrating a paradigm shift through the integration of physical laws such as computational fluid dynamics (CFD) and thermal systems, enhanced data efficiency, and improved interpretability and trustworthiness. By embedding governing physical principles into model architectures, loss functions and optimization routines, PIML not only reduces the solution space but also stabilizes predictions in scenarios with scarce or noisy data (Seyyedi et al., 2023; Sharma et al., 2023). This approach facilitates superior generalization and extrapolation capabilities, enabling accurate inference in complex, high-dimensional multi-physics problems (Karniadakis et al., 2021). Furthermore, rigorous uncertainty quantification is achieved by leveraging physical constraints, thereby bolstering model trustworthiness and reliability (Mao & Jin, 2024). In addition, PIML offers computational efficiency by diminishing the need for extensive datasets and accelerating convergence, while ensuring robust model stability. Finally, comprehensive validation and verification protocols ensure that these models maintain both transparency and interpretability, fostering increased confidence in their deployment for diverse engineering challenges (Xie et al., 2023), thus promoting innovation.

Table 3: The benefits of modern PIML in engineering: Enhancing accuracy, efficiency, and predictive capabilities

| Benefits of PIML | Key Considerations | Engineering Applications | PIML Model | References |
|---|---|---|---|---|
| Integration of physical laws | Embed physics; use hybrid modelling | CFD, structural mechanics, thermal systems | PINNs: Navier–Stokes solver | (Yazdani & Tahani, 2024; Gu et al., 2024; Seyyedi et al., 2023) |
| Data efficiency and scarcity | Reduce data; balance quality and quantity | Structural monitoring, aerospace design | PINNs predict stress from sparse sensors | (Cheruku et al., 2025; Mao & Jin, 2024; Hao et al., 2023; Sharma et al., 2023) |
| Interpretability and trustworthiness | Ensure physics consistency; boost explainability | Predictive maintenance, robotics, process control | Hybrid model for interpretable diagnostics | (Ali et al., 2024; Y. Wu et al., 2024; Seyyedi et al., 2023) |
| Generalisation and extrapolation | Robust across regimes; transferable modelling | Multi-physics, environmental, energy systems | Optimizable PINNs for varying boundaries | (Nyangon, 2025a; Torchio et al., 2025; Pateras et al., 2023) |
| Uncertainty quantification | Propagate errors; integrate probabilistic measures | Risk assessment, safety-critical simulations | Bayesian PINNs provide uncertainty bounds | (Huber et al., 2023; Ma et al., 2022; Molnar & Grauer, 2022) |
| Computational efficiency | Balance physics solvers and data efficiency | Real-time monitoring, control engineering | Reduced-order PINNs enable fast real-time simulation | (Cheruku et al., 2025; Rohrhofer et al., 2023; Xie et al., 2023) |

| Model stability and convergence | Ensure numerical stability; optimize convergence | Stability analysis, vibration studies, circuit simulations | Adaptive PINNs improve training convergence | (Gu et al., 2024; Joshi et al., 2024; Yin et al., 2024) |
|---|---|---|---|---|
| Validation and verification | Benchmark rigorously; perform systematic error analysis | Calibration in automotive, aerospace, energy systems | PIML cross-validated against experimental data | (Nyangon, 2025a; Diz et al., 2023; Kustowski et al., 2022; Carleo et al., 2019) |

## 3. Applications of PIML in Electrical Machines and Drives

### 3.1 Field Analysis and Simulation

In electrical machines and drives, physics consistency is fundamental to ensuring that electromagnetic field simulations accurately reflect real-world behavior. Central to this consistency are Maxwell's equations—Gauss's law for electricity, Gauss's law for magnetism, Faraday's law of induction, and Ampère's law with Maxwell's correction—which together describe how electric and magnetic fields are generated and interact with charges and currents (Watthewaduge et al., 2020). To compute the electromagnetic torque in an electrical machine, one typically begins with the full set of Maxwell's equations and then uses them to derive the forces and moments acting on the machine's moving parts. The key steps involve:

(a) Maxwell's equations and constitutive relations:

The differential form of Maxwell's equations are given by:

$$\nabla \cdot \mathbf{D} = \rho \qquad \text{(Gauss's law for electricity)}$$
$$\nabla \cdot \mathbf{B} = 0 \qquad \text{(Gauss's law for magnetism)}$$
$$\nabla \times \mathbf{E} = -\frac{\partial \mathbf{B}}{\partial t} \qquad \text{(Faraday's law)}$$
$$\nabla \times \mathbf{H} = \mathbf{J} + \frac{\partial \mathbf{D}}{\partial t} \qquad \text{(Ampère's law with Maxwell's correction)}$$

The fields are linked by the constitutive relations, which in a linear medium are:

$$\mathbf{D} = \varepsilon \mathbf{E}, \quad \mathbf{B} = \mu \mathbf{H},$$

where,

- $\varepsilon$ is the permittivity,
- $\mu$ is the permeability,
- $\rho$ is the charge density, and
- J is the current density.

(b) Force and torque density:

The Lorentz force density, representing the force per unit volume, is expressed as:

$$\mathrm{f} = \rho\,\mathbf{E} + \mathbf{J} \times \mathbf{B}.$$

The electromagnetic torque T acting on a volume $V$ can then be calculated by taking the moment of the force density:

$$\mathbf{T} = \int_{\mathrm{V}} \mathbf{r} \times \mathbf{f}\,\mathrm{dV},$$

where r is the position vector relative to the axis of rotation.

(c) Maxwell stress tensor approach:

A more general and often more convenient formulation is to express the torque using the Maxwell stress tensor. The Maxwell stress tensor $\mathrm{T}_{\mathrm{stress}}$ for a medium with permittivity $\varepsilon$ and permeability $\mu$ is defined as:

$$\mathbf{T}_{\mathrm{stress}} = \mathbf{DE} + \mathbf{BH} - \frac{1}{2}(\mathbf{E} \cdot \mathbf{D} + \mathrm{B} \cdot \mathbf{H})\mathrm{I},$$

where I is the identity tensor. Here, the dyadic products DE and BH represent second-order tensors formed from the field vectors.

The torque can then be computed by integrating the moment due to the stress tensor over a closed surface $S$ that encloses the volume of interest:

$$\mathbf{T} = \oint_{\mathrm{S}} \mathbf{r} \times (\mathbf{T}_{\mathrm{stress}} \cdot \mathbf{n})\mathrm{dS},$$

where $n$ is the outward unit normal to the surface S.

(d) Application in electrical machines: In electrical machines, such as motors or drives, the interaction between the stator and rotor plays a critical role in their operation.

- Stator and rotor interaction: The stator windings generate a time-varying magnetic field, as described by Ampère's law. This field then interacts with the currents induced in the rotor, or with pre-existing rotor currents in the case of synchronous machines, generating a Lorentz force.
- Torque production: The distribution of electromagnetic fields, which is determined by Maxwell's equations, dictates the force density within the machine. By integrating the moments of these forces (or using the Maxwell stress tensor method) over the geometry of the machine, the net electromagnetic torque that drives the rotor can be computed.

This formulation links the fundamental Maxwell's equations directly to the torque computation, facilitating the accurate modelling and simulation of complex electromagnetic devices. Combining these elements, a detailed expression for the electromagnetic torque $T$ in an electrical machine can be written as:

$$T = \oint_S r \times \left\{ \left[ DE + BH - \frac{1}{2}(E \cdot D + B \cdot H) I \right] \cdot n \right\} dS.$$

This integral formulation can be used to compute the torque based on the distribution of electric and magnetic fields, which are determined by solving Maxwell's equations with appropriate boundary conditions and material properties within the machine. These equations provide the theoretical foundation for calculating electromagnetic torque in electrical machines and drives. Additionally, they inform the selection of boundary conditions, such as Dirichlet and Neumann types, which govern the field behavior at material interfaces. Advanced numerical methods like the finite element method (FEM) and finite difference method (FDM) are applied to discretize the domain and solve these equations, thereby preserving both local field details and global performance characteristics (Roubache et al., 2018). This rigorous application of theory ensures that the simulation honors fundamental physical laws, enabling accurate predictions across varied operating conditions. In addition, cross-validation against analytical solutions and experimental measurements reinforces the robustness of these models, while continual refinements in solver algorithms further enhance computational reliability.

Data integration in field analysis requires leveraging sparse experimental or sensor data to calibrate and validate simulation models effectively. Techniques such as data assimilation and transfer learning enable the integration of limited sensor readings with high-fidelity numerical simulations, thus bridging the gap between theoretical predictions and practical measurements (Alatawneh & Pillay, 2016). By incorporating sparse data into the calibration process, researchers can refine simulation outputs and adjust model parameters, which in turn improves the overall accuracy of field predictions. Simultaneously, accounting for material nonlinearity—such as magnetic saturation and hysteresis effects—is crucial for capturing the true behavior of electrical machines. Advanced control strategies and iterative methods, including feedback linearization and refined current sheet models, have been developed to address these nonlinearities (Accetta et al., 2022; J. Guo et al., 2020). These approaches not only capture the inherent nonlinearity of magnetic materials but also mitigate instability in simulations. Furthermore, by integrating sensor calibration results with nonlinear solvers, engineers can enhance model fidelity, ensuring that both local material behaviors and broader field interactions are accurately represented. This synthesis of sparse data and nonlinear modelling is vital for developing reliable simulations that can support advanced machine design and operational optimization.

Achieving computational efficiency without sacrificing simulation fidelity is a constant challenge in field analysis for electrical machines and drives. High-fidelity simulations, while essential for capturing intricate electromagnetic interactions, must be balanced with real-time or iterative design requirements. Surrogate modelling techniques, mesh optimization strategies, and hardware acceleration (via FPGAs or GPUs) have emerged as effective solutions to reduce computational overhead (Mojlish et al., 2017; Praslicka et al., 2023; Tahkola et al., 2020). These methods optimize mesh resolution and solver performance, allowing detailed simulations to be performed more rapidly. Equally important is the incorporation of error bounds and probabilistic measures to assess the reliability of simulation predictions. Techniques such as Bayesian calibration, Monte Carlo methods, and Gaussian process regression provide quantitative estimates of uncertainty, enabling engineers to gauge the confidence of their results (Ma et al., 2022; Manfredi & Trinchero, 2022). By combining high-fidelity physics-based models with data-driven uncertainty quantification frameworks, researchers can deliver robust field predictions that inform both design optimization and risk assessment. Ultimately, this integrated approach supports the development of electrical machines and drives that are not only high performing but also resilient to variabilities in material properties and operational conditions.

### 3.1.1 Utilizing PINNs to Model Complex Electromagnetic Phenomena

PINNs offer a transformative framework for modelling complex electromagnetic phenomena in electrical machines and drives by embedding relevant physical constraints. To maintain physical fidelity in field predictions, it is imperative that PINNs incorporate governing equations—such as Maxwell's equations—and appropriate boundary conditions directly within the loss function (Bajaj et al., 2023). This integration ensures that the network honors fundamental conservation laws and electromagnetic principles, thereby reducing the occurrence of non-physical solutions that often plague purely data-driven models. By aligning the neural network's outputs with established analytical benchmarks and experimental data, PINNs achieve enhanced reliability in simulating field distributions, thermal effects, and other intricate behaviors inherent to electrical drives. In practice, the careful selection and tuning of boundary conditions, together with the incorporation of physical laws, provide an intrinsic regularization that mitigates the adverse effects of sparse or noisy measurements. Furthermore, advances in automatic differentiation have facilitated the efficient computation of complex derivatives, thereby streamlining the training process. This robust integration of physics not only enhances simulation fidelity but also bridges the gap between theoretical constructs and practical engineering applications, ultimately supporting innovative design and control strategies.

Balancing of optimization and computational efficiency represents an equally critical challenge in deploying PINNs for electrical machines and drives. The key lies in harmonizing the data-driven loss function with embedded physics-based constraints to ensure rapid and stable convergence without sacrificing model accuracy (Ghattas & Willcox, 2021). By carefully calibrating the weightings between empirical data and the governing equations, researchers can reduce error propagation and numerical instability, which are particularly problematic in high-dimensional and multi-scale scenarios. Recent innovations—such as adaptive learning rate schedules, gradient regularization, and surrogate modelling techniques like Physics-Informed Gaussian Process (PIGP) regression—have been instrumental in managing computational complexity while maintaining high-fidelity predictions (Huber et al., 2023). Moreover, hybrid frameworks that combine conventional numerical methods with deep learning approaches further alleviate computational burdens and accelerate the design optimization process. These advancements not only enhance convergence rates but also provide deeper insights into the underlying physics, thus fostering robust, scalable simulation models for real-world electrical machine design and control applications.

Data integration is another critical consideration for leveraging PINNs in electrical machines and drives. Field simulations often rely on sparse experimental or sensor data that may not fully capture the underlying electromagnetic behavior. PINNs address this challenge by fusing limited but high-quality observational data with physics-based constraints during the training process. By incorporating boundary measurements, sensor outputs, and calibration data, these networks can adjust internal parameters to correct for systematic discrepancies inherent in simulation-only models. Techniques such as transfer learning and data assimilation have been employed to enhance this hybrid approach, enabling PINNs to generalize across diverse operating conditions (Alatawneh & Pillay, 2016; Parekh et al., 2023). This strategy not only mitigates the effects of data sparsity but also creates a self-correcting model that refines its predictions in real time. By integrating multi-modal data sources into the training workflow, PINNs facilitate a more robust calibration of electromagnetic field predictions, ultimately ensuring that simulations remain representative of practical machine behavior under variable conditions.

Handling material nonlinearity and multi-scale modelling are essential when addressing the complex behaviors encountered in electrical machines and drives. Nonlinear effects such as magnetic saturation and hysteresis challenge traditional numerical methods due to their dynamic and often unpredictable nature. PINNs offer an innovative solution by incorporating differential equations that characterize these nonlinear properties, enabling the network to adaptively learn from iterative feedback during training (Yin et al., 2024; J. Guo et al., 2020). At the same time, multi-scale modelling techniques are integrated to reconcile the fine-scale interactions—such as local eddy current losses and microscopic flux variations—with the broader electromagnetic performance of the entire machine. This coupling of local and global scales is achieved by designing network architectures that capture both detailed microstructural effects and overall system behavior. The combined approach not only enhances the fidelity of the simulations but also provides insights into the interplay between material properties and operational performance across various spatial and temporal dimensions, thereby offering a cohesive strategy for the modelling of complex electromagnetic systems.

Balancing high-fidelity simulation with real-time design requirements and incorporating uncertainty quantification is critical for modern electrical machine design. PINNs address this dual challenge by optimizing computational efficiency through surrogate modelling techniques that reduce reliance on finely discretized meshes and lengthy numerical solvers. By embedding physics-based constraints, these networks facilitate rapid convergence even when operating with coarser discretization, thereby maintaining accuracy while significantly lowering computational costs (Jansson et al., 2020; Praslicka et al., 2023). Moreover, uncertainty quantification is achieved by incorporating error bounds and probabilistic measures directly into the training regimen. Advanced methods, such as Bayesian calibration and Monte Carlo simulations, allow the network to generate confidence intervals for predicted field values, effectively assessing the reliability of simulation outputs (Galetzka et al., 2019; Ma et al., 2022). This integration of uncertainty measures not only enhances the robustness of the simulation outcomes but also facilitates iterative design optimization by clearly identifying areas of potential error. Consequently, PINNs offer a compelling framework that reconciles the need for high-fidelity electromagnetic simulations with the practical constraints of real-time engineering applications.

### 3.1.2 Transfer Learning in Electromagnetic Analysis

Transfer learning in electromagnetic analysis, particularly when employing PINNs for electrical machines and drives, necessitates the rigorous preservation of Maxwell's equations and associated boundary conditions during the transfer process. In this framework, the underlying physical laws are embedded within the network's loss function, ensuring that the simulation-trained model retains its physics consistency during retraining (Bajaj et al., 2023). To maintain fidelity, the transfer learning process incorporates explicit constraints derived from Maxwell's equations, thereby enforcing correct field behavior even when the model is adapted to new data. Moreover, careful integration of boundary conditions – both Dirichlet and Neumann types – ensures that the fundamental electromagnetic interactions are not compromised during parameter refinement. This approach yields a robust initial model that not only captures the nuances of electromagnetic phenomena but also provides a solid baseline for subsequent calibration against experimental measurements, ensuring that the model remains physically plausible while adapting to new operational conditions.

Evaluating the similarities between source and target tasks is essential for adjusting the transferred model to achieve robust performance. In practice, comparative analyses of geometry, material properties, and operating conditions are conducted to determine the degree of overlap between the simulation domain and real-world configurations (Ghattas & Willcox, 2021). This evaluation guides the selection of training data and the adaptation of network architectures to handle variations without incurring overfitting or loss of convergence. Additionally, data utilization is carefully balanced with computational resource management by implementing efficient retraining strategies that combine sparse experimental datasets with large-scale simulation outputs. Such hybrid calibration techniques enable the model to refine its predictions iteratively, ensuring that both convergence and accuracy are maintained while computational complexity is minimized (Soibam et al., 2024).

### 3.2 Performance Estimation

Performance estimation is the process of quantifying the expected loss of a predictive model when applied to unseen data, serving as a critical metric for validating generalization (Brosch et al., 2021). In the context of electrical machines and drives, this process is particularly challenging due to the inherent nonlinearity, temporal dependencies, and uncertainties present in the system dynamics. Researchers have shown that traditional methods may fall short when dealing with complex motor behaviors and transient phenomena (Kirchgässner et al., 2021). As a result, hybrid approaches that combine first-principles modelling with data-driven corrections are increasingly gaining traction. These techniques aim to overcome limitations related to model mismatches and measurement noise, which are often exacerbated by the limited availability of training data. Overall, robust performance estimation not only requires careful selection of estimation techniques but also a nuanced understanding of the underlying physical processes, ensuring that models remain reliable and physically consistent across varying operating conditions.

#### 3.2.1 Hybrid Models for Accurate State Estimation

A critical component in advancing performance estimation for electrical machines is the development of hybrid models for accurate state estimation. These models integrate traditional physics–based representations with modern ML methods to compensate for unmodelled dynamics and uncertainties. By leveraging fundamental physical laws, hybrid models ensure physical consistency while incorporating data–driven corrections to address phenomena such as flux linkage harmonics and inverter nonlinearities (Brosch et al., 2021). This approach necessitates that the model dynamically adapts to deviations from nominal conditions, thereby enhancing its robustness against noise and measurement errors. In addition, the integration of data–driven elements allow for capturing transient behaviors with greater precision, which is essential in systems where rapid changes can lead to instability. Although promising, this strategy poses challenges in maintaining the balance between fidelity to physical laws and the flexibility of ML adjustments, ultimately impacting the overall accuracy of performance estimation in complex drive systems.

### 3.2.2 Online Parameter Identification

In parallel, online parameter identification is pivotal for real–time performance estimation in electrical drives. In these applications, the system must continuously update its state estimates and model parameters as new data become available, ensuring that the model remains responsive to changing operating conditions. Speed and responsiveness are therefore crucial, as delays in adapting to transient changes can lead to inaccurate state estimations. Moreover, online methods must be robust to noise, given that measurement uncertainties and disturbances are inherent in real–world applications (Zhu et al., 2021). Computational efficiency is equally important, as the algorithms employed need to operate within stringent time constraints without compromising accuracy. These methods often rely on recursive or iterative algorithms that adjust parameters in real time, but the challenge remains to balance rapid adaptation with the prevention of error propagation—a common pitfall when the training data are sparse or highly variable (Cerqueira et al., 2020). Consequently, ongoing research in online parameter identification focuses on optimizing these tradeoffs to enhance the overall reliability of ML applications in electrical machine control.

### 3.2.2 Offline Parameter Identification

Offline parameter identification, by contrast, leverages accumulated historical data to refine and calibrate the underlying models over longer time horizons. This approach facilitates deep calibration, allowing for computationally intensive analyses that yield improved long–term accuracy. By periodically recalibrating the model, offline methods can address systematic discrepancies between predicted and actual performance, effectively incorporating error analysis and correction into the estimation process (Wallscheid, 2021). Such recalibration is particularly beneficial in environments where operating conditions evolve slowly, and short–term fluctuations may obscure longer–term trends. Moreover, offline identification supports model refinement by integrating insights derived from comprehensive data sets, which can lead to a more robust representation of the machine's behavior under varied conditions. However, this process is not without challenges; ensuring that the historical data accurately reflect current operating conditions requires careful selection and validation of the training data. As such, while offline parameter identification offers a pathway to enhanced accuracy and reliability, it must be implemented in conjunction with strategies that continuously monitor and adjust for discrepancies over time.

## 3.3 Design and Optimization

### 3.3.1 ML-Driven Design and Optimization Methods with Physical Constraints

Many researchers have identified three pivotal methods for the design and optimization of electrical machines and drives: deterministic optimization (specifically gradient-based optimization), stochastic optimization (metaheuristic approaches), and surrogate modelling using ML and ANN (Asef & Vagg, 2024; Tahkola et al., 2020). Deterministic optimization leverages gradient-based methods to systematically refine design variables by directly incorporating physical constraints into the optimization process. This approach not only enhances convergence towards an optimal solution but also ensures that the resultant model adheres to the underlying physics of the system (Nghiem et al., 2023). In contrast, stochastic optimization utilizes metaheuristic strategies, such as genetic algorithms and physics-informed Bayesian optimization, to explore the design space in a more globalized manner. These methods are particularly useful when the optimization landscape is non-convex and riddled with local minima, thereby enabling the discovery of solutions that might be overlooked by purely gradient-based methods (Antonion et al., 2024; Zhao et al., 2024). Complementing these techniques, surrogate modelling involves the construction of computationally efficient, ML–based approximations that replicate the behavior of more complex physical simulations. By integrating data-driven models with physical constraints, surrogate models provide rapid predictions of system performance, which is crucial when experimental data are scarce or simulation costs are prohibitively high (Tahkola et al., 2020).

The integration of these three methods under the umbrella of PIML not only bridges the gap between theoretical simulation and practical application but also fosters a more robust and physically plausible design framework for electric machines and drives. Deterministic optimization methods are particularly well-suited for scenarios where the design variables are continuous, and the system's behavior can be reliably captured by differentiable models. These methods, by utilizing gradient information, offer a direct path to improvement and are essential for applications requiring fine-tuned performance adjustments (Chen et al., 2023; Joshi et al., 2024). Meanwhile, stochastic optimization provides a valuable complement by introducing randomization and probabilistic exploration into the design process, which is advantageous when the objective function exhibits discontinuities or when the design space is highly complex. Finally, surrogate modelling stands out by significantly reducing computational overhead. When physical experiments or high-fidelity simulations are time-consuming or expensive, ML–based surrogate models enable rapid assessments and iterative improvements, thereby accelerating the overall design cycle. These methods underscore the potential of PIML to yield efficient, accurate, and physically consistent optimization strategies for the next generation of electrical machines and drives (Asef & Vagg, 2024; Lei et al., 2017; Nghiem et al., 2023).

### 3.3.2 Optimization of Magnetic Materials and Geometries

Magnetic materials and geometries are fundamental to the design and performance of electrical machines and drives, as they determine the efficiency, power density, and operational stability of such systems. Magnetic materials, including soft magnets used in transformer cores and hard magnets in motors, are engineered to exhibit properties such as high magnetic flux density, optimal coercivity, and low core losses. Equally, the geometry of magnetic assemblies—encompassing the shape, alignment, and spatial distribution of magnetic elements—plays a critical role in minimizing energy losses and maximizing performance. Recent advances in PIML optimization methods have quantitatively enhanced these properties by integrating multiscale simulations with statistical learning techniques (Wu et al., 2024; Kovacs et al., 2020; Hsieh et al., 2018). For instance, regression models and active learning pipelines have enabled precise predictions of magnetic saturation and coercivity, thereby guiding the tuning of material compositions and process parameters to achieve superior energy efficiency (Wang et al., 2020).

PIML methods quantitatively improve magnetic material characteristics through systematic optimization of both intrinsic and extrinsic properties. By leveraging physics-based constraints alongside data-driven techniques, these methods refine predictive models that quantify the interplay between microstructural features and magnetic performance. For example, Bayesian optimization and transfer learning have been applied to fine-tune process variables, resulting in enhanced coercivity and increased flux density—crucial for applications in electric vehicle drives and high-efficiency motors (Lambard et al., 2022; Fukushige et al., 2013). This approach not only reduces computational overhead by streamlining the simulation of multiple data modalities but also constrains the solution space, leading to robust models that generalize well despite limited experimental data. Ultimately, the integration of PIML in magnetic material design provides a quantitative framework for optimizing energy efficiency and performance, thus paving the way for the development of advanced electrical machines with reduced reliance on rare-earth elements. Figure 6 illustrates PIML applications in electrical machines, including field simulation, performance estimation, design optimization, neural control, digital-twin calibration, and deployment pipeline.

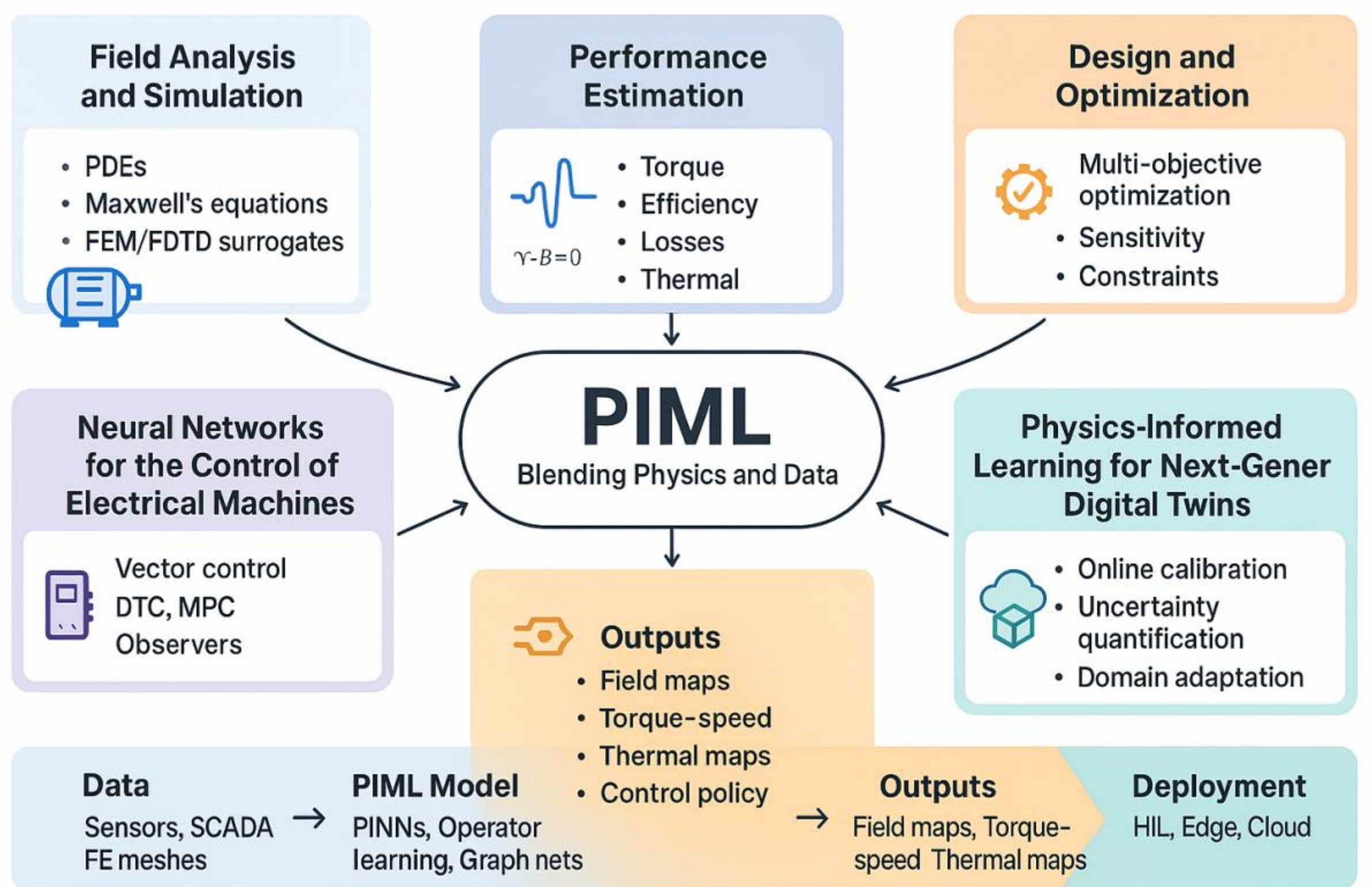

Figure 6: PIML for electrical machines and drives: performance, control, optimization and digital twins

### 3.4 Neural Networks for the Control of Electrical Machines

### 3.4.1 Intelligent Control Strategies for Drives

PIML models have been used in model-based control methods, including Model Predictive Control (MPC), which represents a paradigm that integrates detailed physics-based models directly within an optimization framework to predict future system behavior and manage constraints in electric drives. This approach employs explicit dynamic models to forecast system performance over a receding horizon, thereby enabling real-time adjustment of control actions (Wang et al., 2017). By incorporating ML techniques, particularly those informed by underlying physical principles, PIML-MPC enhances the accuracy of predictions even under conditions of parameter uncertainty and external disturbances (Xue et al., 2023). Recent advancements have addressed computational challenges and improved robustness through innovative weighting factor calculations and harmonic distortion optimization (Elmorshedy et al., 2021). The ability to embed comprehensive models within the predictive framework not only leverages the strengths of both physics-based and data-driven approaches but also paves the way for more reliable control in complex electric drive systems (Wang et al., 2017).

Reinforcement Learning (RL)–based control constitutes a second key strategy, utilizing advanced RL algorithms to iteratively learn optimal control policies through direct interaction with the drive system. This method circumvents the need for an explicit system model by adapting to nonlinearities and uncertainties via experience-driven improvements (Kazemikia, 2024; Farah et al., 2023). When augmented with physics-informed constraints, these algorithms can provide enhanced robustness and adaptability, effectively balancing exploration and exploitation in dynamic operating environments (Traue et al., 2022; Book et al., 2021). Moreover, RL-based controllers have demonstrated promising results in simulation and preliminary experimental settings, particularly for motor types such as permanent magnet synchronous and DC motors, where conventional controllers often struggle (Schenke et al., 2020). While computational complexity and issues with sim-to-real transfer remain challenging, ongoing research continues to refine these methods, thus progressively positioning RL-based control as a viable solution for next-generation electric drive applications (Kazemikia, 2024).

Thirdly, hybrid intelligent control, which synergistically combines Adaptive Neuro-Fuzzy Inference Systems (ANFIS) with data-driven methodologies, is emerging as a complementary strategy that fuses the strengths of neural networks, fuzzy logic, and classical control techniques. This strategy capitalizes on the capability of ANFIS to approximate nonlinear relationships and adapt to changing system dynamics, thereby ensuring improved parameter adaptation and fault tolerance in electric drives (Gupta et al., 2024; Parra et al., 2018). By merging traditional model knowledge with ML insights, hybrid intelligent control not only mitigates the limitations inherent in singular approaches but also offers significant improvements in energy management, dynamic performance, and overall system efficiency (Rajasekar & Ashok, 2024; Nyangon, 2021a; Suhail et al., 2021). Recent studies have showcased the superiority of these hybrid systems in applications ranging from torque vectoring and regenerative braking to real-time optimization in multi-motor setups (Gaber et al., 2021). Despite the challenges of integrating diverse methodologies, the hybrid approach represents a robust and flexible framework that is well suited to the demands of modern electric drive systems (Parra et al., 2018).

### 3.4.2 Adaptive Controllers Based on PIML Models

Adaptive controllers are control systems that dynamically adjust their parameters in response to real-time performance and environmental uncertainties, thereby compensating for model discrepancies, external disturbances, and time-varying dynamics (Nyangon, 2025a; Nghiem et al., 2023; Pereida et al., 2019). Recent advancements in PIML have invigorated this domain by integrating physical constraints into neural network frameworks, ultimately enabling a more accurate and robust modelling of complex systems. In the context of electrical machines and drives, these PIML-based adaptive controllers harness the inherent ability of neural networks to approximate highly nonlinear functions while embedding the governing physical laws into the training process (Hanover et al., 2022; Yousef et al., 2019). Such an approach significantly enhances the controller's capacity to stabilize systems under variable load conditions and unpredictable disturbances, thereby improving precision in trajectory tracking and overall performance. Moreover, the integration of PIML techniques with established control methods—such as iterative learning control (ILC) and MPC—further reinforces system robustness by compensating for systematic tracking errors and model uncertainties (Staessens et al., 2022; Pereida et al., 2019).

In practical applications, especially within the realm of electrical machines and drives, PIML-based adaptive controllers have demonstrated a marked improvement in both stability and precision. For instance, by leveraging a combination of neural-network-assisted control and physics-based constraints, these controllers can mitigate the adverse effects of nonlinearities, hysteresis, and measurement noise that are typically prevalent in electrical drive systems (Feng et al., 2020; Yang et al., 2019). The adaptability conferred by such systems not only permits a rapid convergence towards the desired operating regime but also ensures that error propagation remains controlled even when confronted with imperfect boundary conditions or varying operational demands ((Nan & Hutter, 2024; Nghiem et al., 2023). This dual capability of maintaining robust performance while facilitating transfer learning across distinct dynamic environments—ranging from robotic motion control to high-precision tracking in piezoelectric actuators—underscores the potential of PIML-based adaptive controllers as industrial-grade solutions (Hanover et al., 2022; Yousef et al., 2019). Generally, the enhanced stability and precision afforded by these controllers, as evidenced by both simulation studies and experimental validations, highlight their suitability for complex, safety-critical applications, thereby reinforcing their emerging role as a pivotal tool in advanced control system design (Staessens et al., 2022; Pereida et al., 2019).

## 3.5 Physics-Informed Learning for Next-Generation Digital Twins

Digital twins are virtual replicas of physical systems that replicate real-time behavior, enabling precise simulation and continuous monitoring of electrical machines and drives (Bouzid et al., 2020). These digital twins can be integrated with existing control systems by adopting a systematic four-part modelling process, as presented by Liu et al. (2022), that includes: the physical entity, a perception layer for comprehensive data acquisition, a middle layer for data processing and simulation, and a decision layer for human–computer interaction (Figure 7). The perception layer is tasked with gathering extensive operational data—ranging from motion drive metrics to status signals—while the middle layer synthesizes this information into robust simulation models using multidisciplinary design fundamentals, such as finite element analysis and data-driven approaches (Liu et al., 2022).

The decision layer then maps the digital data to actionable insights, thereby facilitating closed-loop feedback and timely intervention by control systems. Recent studies have shown that the integration of digital twins with existing controllers—exemplified by their application in power electronic converters and rotating machinery—significantly enhances real-time diagnostics and fault prediction (Nyangon, 2025b; Cheruku et al., 2025; Milton et al., 2020). Moreover, when digital twins are embedded within Manufacturing Execution Systems (MES), they trigger automated actions on physical equipment, ensuring synchronized and accurate operation between digital simulations and real-world processes (Nyangon, 2025b; Negri et al., 2020). This integrated approach not only improves the precision of real-time monitoring but also mitigates the propagation of uncertainties arising from sensor inaccuracies, thereby reinforcing the overall stability and resilience of control systems (Cheruku et al., 2025).

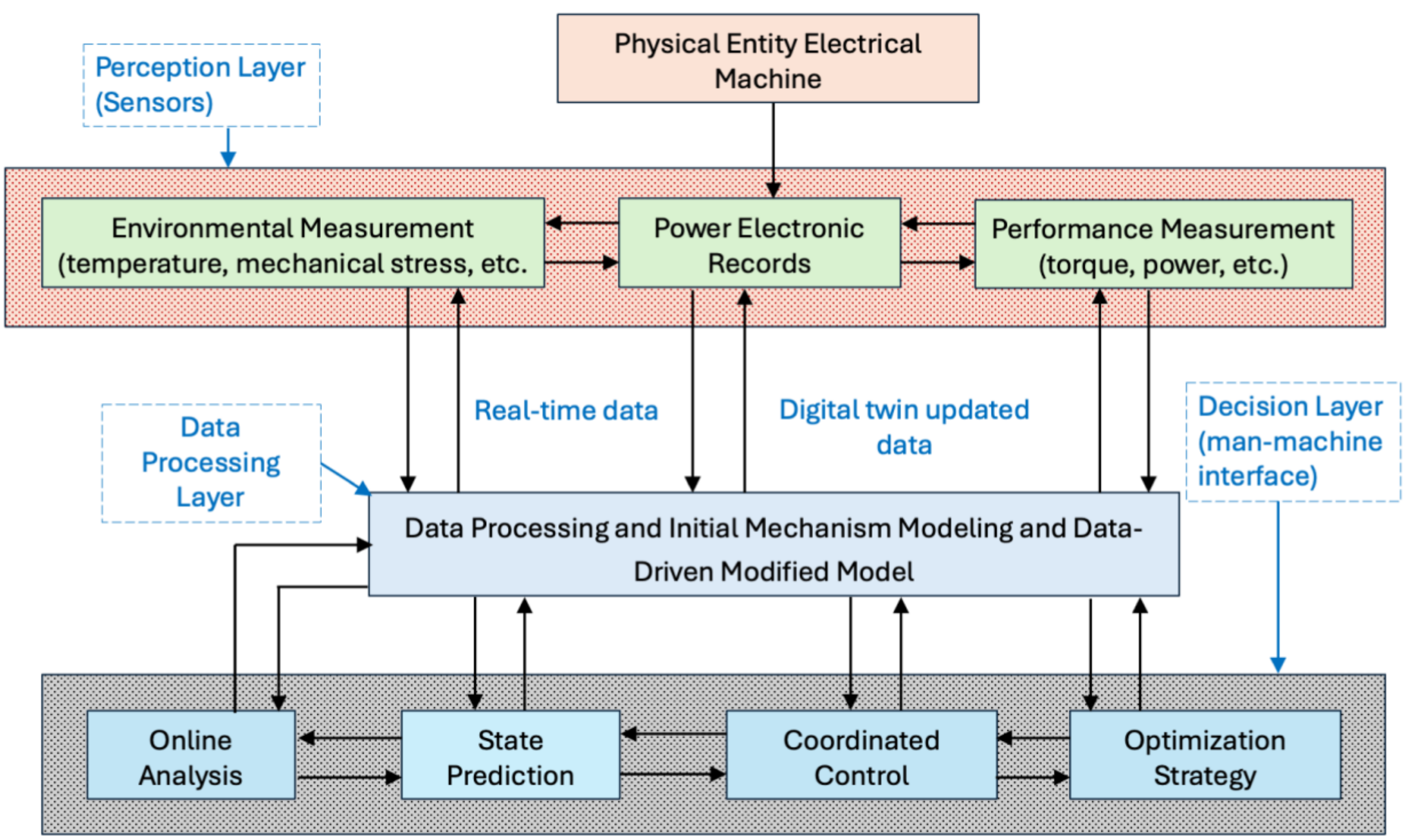

Figure 7: The four key components of the digital twin modelling framework for electrical drive systems

### 3.5.1 Real-Time Monitoring and Diagnostics

Real-time monitoring and diagnostics, underpinned by PIML, offer a transformative approach to the management of electrical machines and drives. Digital twins serve as high-fidelity virtual replicas, enabling dynamic simulation and continuous evaluation of complex electromagnetic phenomena and thermal behaviors inherent in electrical systems (Cheruku et al., 2025; Liu et al., 2022). By integrating data-driven models with traditional finite element methods and probabilistic frameworks, these systems facilitate fault detection, design optimization, and predictive maintenance. The convergence of PIML and digital twin technology allows for rapid, precise characterization of machine states, even when physical measurements are challenging or incomplete. Such capabilities are crucial for maintaining system reliability and ensuring operational efficiency, particularly in high-demand industrial environments where timely interventions can avert catastrophic failures (Milton et al., 2020). Furthermore, the incorporation of real-time data into these models supports continuous calibration and adjustment, thereby enhancing both the accuracy and robustness of diagnostics across a range of electrical drive applications. These integrated methodologies ensure that diagnostic outputs are not only precise but also resilient to the uncertainties inherent in sensor measurements and environmental variations.

Digital twins are deployed to support real-time monitoring and diagnostics by leveraging advanced sensor integration and sophisticated control algorithms. In practical applications, digital twin frameworks have been embedded within controller architectures, enabling a closed-loop feedback mechanism that synchronizes virtual simulations with the physical behavior of electrical machines (Liu et al., 2022). Recent studies have demonstrated that by utilizing combinatorial data synthesis strategies, digital twins can improve the predictive performance of reduced-order models, thereby offering higher fault detection accuracy (Cheruku et al., 2025; Nyangon, 2025b). The use of augmented reality tools further facilitates interactive human intervention, as real-time alerts and visualizations assist operators in diagnosing and rectifying anomalies promptly. Additionally, the partitioning of digital twin models into discrete control layers permits targeted analysis of power electronic converters and rotating machinery, ensuring that both global system behavior and local fault conditions are comprehensively monitored (Bouzid et al., 2020; Diz et al., 2023). This multi-layered approach not only mitigates the propagation of measurement errors but also enhances the scalability of digital twin applications across diverse industrial scenarios. Consequently, the synergy between digital twin technology and PIML emerges as a potent enabler for intelligent, real-time diagnostics and sustainable management of electrical machines and drives (Cheruku et al., 2025; Nyangon, 2025a; Falekas & Karlis, 2021). Moreover, ongoing research continues to refine these digital twin frameworks, addressing challenges such as data synthesis variability and computational efficiency to further enhance system robustness and real-time performance. These advancements promise substantial industrial impact.

### 3.5.2. Comparative Analysis of Traditional ML and PIML Approaches

Traditional ML methods, such as linear regression, logistic regression, decision trees, random forests, and support vector machines (SVMs), have long served as valuable tools in enhancing electrical machine and drive performance. These techniques primarily rely on statistical inference and data-driven optimization, utilizing operational data to identify and model relationships between input variables and system behavior. Linear and logistic regression provide computational simplicity and ease of interpretation, while decision trees and random forests capture non-linear interactions with relative efficiency (Gawande, 2024). SVMs, in particular, offer robust performance in handling complex decision boundaries, enabling precise fault diagnosis and predictive control (Wu et al., 2024). Nevertheless, these traditional methods are inherently limited by their inability to incorporate explicit physical laws governing the underlying system dynamics.

Their performance often diminishes when confronted with the intricacies of high-dimensional, non-linear processes typical in modern electrical drives, where data may be scarce or noisy. As a result, while conventional approaches deliver rapid implementation and interpretability, they may fall short in scenarios demanding high fidelity and resilience to uncertainty. Thus, these classical techniques invariably require complementary strategies that integrate empirical data with physical models for enhanced real-world performance.

PIML methods, including PINNs, PIGNNs, Deep Operator Networks (DeepONets), FNOs, and PIGP, represent a paradigm shift in addressing the challenges of electrical machine and drive performance. By embedding physical constraints and first-principles knowledge directly into the learning process, these techniques deliver enhanced accuracy and robustness, particularly in systems exhibiting complex, non-linear dynamics (S. Zhao et al., 2022). Table 4 compares key differences between traditional and physics-informed methods for electrical machines and drives applications. PINNs and PGNNs facilitate the integration of differential equation solvers with deep learning, ensuring that the resulting models adhere to underlying physical laws, while DeepONets and FNOs excel at learning operators that map infinite-dimensional spaces with improved generalization (Sharma et al., 2023). Moreover, PIGPs provide a probabilistic framework for uncertainty quantification and compensates for sparse data scenarios. This hybrid approach effectively mitigates the limitations of traditional methods by reducing error propagation and ensuring model consistency across varying operating conditions (Zideh et al., 2024; Huber et al., 2023; Parekh et al., 2023). Consequently, PIML methods have demonstrated superior performance in capturing transient behaviors and accommodating measurement uncertainties, thereby offering a robust alternative for real-time control and fault diagnostics in advanced electrical drive systems. These advancements underscore importance of merging physics with data-driven approaches.

Table 4: Comparative analysis of traditional ML vs. PIML methods for electrical machines drives

| Method | Description | Key feature and strength | Application to electrical machines drives | Method type and citations |
|---|---|---|---|---|
| Linear regression | Models linear relationships between input features and a continuous output. | ▪ Simplicity and high interpretability<br>▪ Low computational cost<br>▪ Effective for linearly-behaving datasets | ▪ Predicting performance metrics (e.g., torque, speed, consumption)<br>▪ Parameter estimation and calibration | Traditional ML Refs: Montgomery et al. (2012); James et al. (2013) |
| Logistic regression | Uses a logistic function to model binary or multi-class outcomes. | ▪ Straightforward implementation<br>▪ Probabilistic outputs<br>▪ Computationally efficient | ▪ Fault detection and classification<br>▪ Operational state monitoring | Traditional ML Refs: Hosmer & Lemeshow (2000); James et al. (2013) |
| Decision trees | Tree-based models that split data based | ▪ High interpretability | ▪ Diagnostics and failure mode detection | Traditional ML Refs: Quinlan |
| | on feature thresholds to make predictions. | ▪ Captures non-linear relationships<br>▪ Minimal data pre-processing required | ▪ Performance categorization | (1986); Breiman et al. (1984) |
| Random forests | An ensemble of decision trees that aggregates predictions to improve accuracy. | ▪ Robustness against overfitting<br>▪ Effective with high-dimensional data<br>▪ Improved predictive accuracy | ▪ Predictive maintenance<br>▪ Reliability assessment<br>▪ Optimization of machine performance | Traditional ML<br>Ref: Breiman (2001) |
| Support vector machines (SVMs) | Finds the optimal hyperplane for classification or regression tasks. | ▪ Effective in high-dimensional spaces<br>▪ Versatile with kernel functions<br>▪ Robust to outliers | ▪ Fault classification<br>▪ Anomaly detection<br>▪ Control system tuning | Traditional ML<br>Ref: Cortes & Vapnik (1995) |
| PINNs | Neural networks that incorporate physical laws (e.g., differential equations) into loss function during training. | ▪ Enforces physical consistency<br>▪ Enhanced optimization and extrapolation<br>▪ Effective with sparse data | ▪ Modelling electromagnetic fields<br>▪ Thermal analysis<br>▪ Dynamic behavior simulation | PIML<br>Ref: Raissi, Perdikaris & Karniadakis (2019) |
| PGNNs | Neural networks that integrate physics-based constraints during training for improved model behavior. | ▪ Balances data-driven learning with physics insights<br>▪ Improved interpretability<br>▪ Better extrapolation capabilities | ▪ Real-time monitoring<br>▪ System control and performance optimization<br>▪ Adaptive behavior prediction | PIML<br>Ref: Karniadakis et al. (2021) |
| Deep Operator Networks (DeepONets) | Networks designed to learn operators that map between infinite-dimensional function spaces. | ▪ Learns complex system operators<br>▪ Optimizes to unseen conditions<br>▪ Operates at an operator level rather than | ▪ Control operator design<br>▪ Dynamic simulation<br>▪ Modelling non-linear behaviors | PIML<br>Ref: Lu et al. (2021) |

| | | pointwise predictions | | |
|---|---|---|---|---|
| Fourier Neural Operators (FNOs) | Neural operators that leverage Fourier transform techniques to learn mappings in function spaces efficiently. | ▪ Efficient handling of high-dimensional data<br>▪ Fast training times<br>▪ Captures long-range dependencies in spatial data | ▪ Fast simulation of spatially distributed phenomena<br>▪ Real-time control applications | PIML<br>Ref: Li et al. (2020) |
| Physics-informed Gaussian processes | Gaussian process models enhanced with physics-based priors to improve predictions and quantify uncertainty. | ▪ Probabilistic predictions with built-in uncertainty quantification<br>▪ Integrates physical constraints for improved robustness<br>▪ Flexible non-parametric modelling. | ▪ Fault detection with uncertainty analysis<br>▪ Reliability assessments<br>▪ Predictive maintenance in environments with limited data | PIML<br>Refs: Solin & Särkkä (2014); related recent studies on physics-informed GPs |

## 4. Challenges and Future Perspectives

Industry 4.0 is transforming the manufacturing landscape by integrating digital technologies, including the Internet of Things (IoT), digital twins, and advanced data analytics, into industrial systems (Nyangon, 2025b; Ji & Abdoli, 2023). In the electrical machines and drives market, this paradigm shift facilitates real-time monitoring, enhanced diagnostic precision, and predictive maintenance through the convergence of physical and digital realms. In this regard, PIML embeds physical laws into data-driven models, thus enhancing model reliability and interpretability. However, the implementation of PIML methods in this context is fraught with challenges. Chief among these is the computational complexity associated with modelling high-dimensional, nonlinear phenomena in electrical systems, which often necessitates sophisticated numerical solvers and extensive computational resources (Achouch et al., 2022). Moreover, the variability and uncertainty in operational data, coupled with context-dependent calibration issues, impede the development of universally robust models. The integration of theoretical models with industrial-scale applications further compounds these difficulties, as the limited availability of high-fidelity training data and the sensitivity of algorithms to noisy inputs often result in unstable predictions (Dalzochio et al., 2020). These issues hinder industrial adoption.

Future expectations for PIML in the electrical machines and drives sector focus on overcoming computational and practical challenges through algorithmic efficiency and research efforts.

Advances in numerical methods and parallel computing in the era of Industry 4.0 are expected to mitigate the computational complexity and scalability issues that currently constrain high-dimensional, nonlinear modelling (Çınar et al., 2020). Moreover, a critical requirement remains the translation of theoretical models into industrial applications, where iterative validation against real-world data is essential to address uncertainties and data imperfections (Ji & Abdoli, 2023; Meriem et al., 2023). This process involves bridging the gap between academic innovation and practical implementation, thereby refining PIML frameworks for reliable performance in dynamic operational environments. Equally important is the opportunity for cross-disciplinary collaboration, as efforts among experts in control engineering, computational physics, and data analytics can drive the creation of resilient and adaptable models (Zonta et al., 2020). These improvements are poised to reinforce the transformative impact of Industry 4.0 on predictive maintenance, system optimization, dynamic behavior simulation, real-time monitoring, control operator design, modelling non-linear behaviors, fault detection with uncertainty analysis, and reliability assessments. With sustained research investment and coordinated efforts to resolve theoretical advances with empirical challenges, PIML is anticipated to offer significant gains in efficiency, reliability, and diagnostic precision for electrical machines and drives in the Industry 4.0 era.

### 4.1 Computational Complexity and Scalability

In the context of Industry 4.0, PIML is set to transform the design and control of electrical machines and drives by dramatically reducing computational complexity. Recent studies have demonstrated that integrating physical principles directly into ML frameworks not only enhances prediction accuracy but also leads to substantial reductions in processing times (Schüller et al., 2024). For instance, the use of physics-informed Bayesian optimization coupled with a maximum entropy sampling algorithm (PIBO-MESA) has yielded processing time reductions of approximately 45% compared with stochastic methods such as Non-Dominated Sorting Genetic Algorithm II (NSGA-II) (Asef & Vagg, 2024; Al-Aghbari & Gujarathi, 2022). Similarly, employing a multi-branch deep neural network has reduced simulation times from several hours—typically in the range of three to five hours—to roughly 100 milliseconds per sample when replacing traditional finite element simulations (Parekh et al., 2023). In an Industry 4.0 framework, where rapid adaptability and real-time monitoring are essential, such improvements in computational efficiency are critical for both design optimization and operational control.

Beyond computational speed, PIML methods substantially contribute to scalability in complex industrial applications. The ability to handle large datasets and complex, high-dimensional models is inherent in many PIML architectures, which combine physics-based constraints with the adaptive learning capabilities of neural networks. This dual approach not only mitigates the burden of traditional simulation techniques but also ensures that models remain robust when scaling from laboratory prototypes to full industrial systems. Integration with digital twin technologies further exemplifies this scalability. A digital twin of an electrical machine or drive assets can incorporate PIML algorithms to simulate and predict the machine or drive behavior in real time, enabling predictive maintenance and adaptive control strategies across large-scale production systems (Cheruku et al., 2025). Such systems are designed to operate seamlessly within cyber-physical frameworks, ensuring that improvements in computational performance directly translate into enhanced system reliability and operational efficiency. As these models are refined, they promise to underpin scalable solutions that are adaptable to a wide range of applications, from traction motor design in electric vehicles to power electronic converter optimization.

Despite these advances, several challenges and future expectations remain regarding computational complexity and scalability. One significant challenge is the inherent trade-off between model complexity and computational efficiency. While deeper and more sophisticated PIML models can capture intricate physical phenomena more accurately, they may also demand increased computational resources. This raises concerns about the generalization of such models when applied to very large or complex systems. Moreover, current studies often rely on specific baseline comparisons, and there is a pressing need for standardized benchmarking procedures that can reliably assess scalability across diverse applications. Future research should focus on testing PIML approaches on larger and more heterogeneous datasets, ensuring that performance improvements observed in controlled experiments can be replicated in full-scale industrial environments. Additionally, further integration of PIML with digital twin platforms is anticipated to facilitate comprehensive, real-time monitoring and control, thus bolstering the overall efficiency and resilience of electrical machine systems in Industry 4.0 (Zhao et al., 2022). More prosaically, while PIML offers promising improvements in computational efficiency and scalability, addressing these challenges will be crucial to fully realize its potential in transforming industrial practices.

### 4.2 Bridging the Gap Between Theory and Industrial Applications

PIML is an emerging pivotal method for bridging the longstanding gap between theoretical models and industrial applications in electrical machines and drives within the Industry 4.0 framework. By integrating rigorous physical laws directly into data-driven models, PIML enables the development of more reliable and computationally efficient tools that are capable of real-time operation. Current theoretical approaches—such as finite element analysis (FEA) and finite difference methods (FDM)—are commonly employed to simulate electromagnetic fields and thermal behavior, as well as discretize and solve partial differential equations governing machine operation. However, these methods often suffer from prohibitive computational times and a lack of robustness when confronted with real-world uncertainties—limitations that hinder their practical deployment. For example, conventional finite element simulations, while highly accurate, are computationally intensive and poorly suited for dynamic industrial environments, whereas PIML methods can drastically reduce simulation times and incorporate uncertainties in boundary conditions and material properties. This fusion of physics-based reasoning with adaptive ML not only enhances predictive accuracy and reliability but also facilitates the seamless integration of digital twins into existing control systems, thereby supporting scalable and adaptive maintenance strategies. Moreover, hybrid models that couple physical constraints with statistical learning are showing promise in improving model generalization across diverse operational scenarios. As noted by Liu et al. (2022) and Parekh et al. (2023), such frameworks offer a viable pathway to overcome the disconnect between laboratory-based theoretical advancements and the practical demands of industrial-scale applications, paving the way for optimized design and robust, real-time control in the rapidly evolving industrial landscape.

### 4.3 Future Directions for Integrating PIML in Electrical Systems

The following opportunities for cross-disciplinary collaboration in PIML are poised to drive innovation by integrating diverse expertise and methodologies to overcome inherent challenges identified in 4.0, namely:

1. Integrating domain expertise—through collaboration between ML experts and specialists in physics, engineering, and mathematics enables the embedding of fundamental physical laws directly into learning models, thereby ensuring realistic predictions and operational consistency across applications in fluid mechanics, chemical engineering, and additive manufacturing. This integration enhances model reliability and interpretability, fostering innovative solutions that are both computationally cost efficient and physically grounded (Wu et al., 2024; Sharma et al., 2023).
2. Joint development of hybrid models—achieved by combining classical simulation techniques with data-driven ML approaches—will empower researchers to leverage the strengths of both paradigms. This synthesis promotes the formulation of modular, stable, and accurate models that effectively capture complex phenomena, thereby overcoming limitations inherent in isolated methodologies and facilitating advanced system identification and control through interdisciplinary insights (Gupta et al., 2024; Nagao et al., 2024; Seyyedi et al., 2023; Hanover et al., 2022).
3. Enhanced data sharing and standardization initiatives—provide another critical avenue for innovation in PIML. By creating shared databases, benchmarks, and experimental protocols, collaborative teams can foster reproducibility and accelerate the systematic validation of PIML methodologies across diverse applications (Torchio et al., 2025; Cerqueira et al., 2020). Such interdisciplinary efforts mitigate challenges related to data structure, security, and interoperability while upholding ethical and legal standards, ultimately paving the way for robust and scalable computational frameworks.
4. Innovative algorithm development—driven by a synergistic fusion of computer science, applied mathematics, and domain-specific expertise—to yield novel computational strategies that effectively address the challenges posed by noisy, sparse, and high-dimensional data (Moradi et al., 2023; Kustowski et al., 2022; Arzani et al., 2021) These cross-disciplinary endeavors foster breakthroughs in optimization, convergence, and scalability, enhancing predictive performance and interpretability and thereby driving transformative advancements in PIML applications.
5. Improved uncertainty quantification and interpretability – realized through close collaboration among statisticians, ML researchers, and domain experts. This collective approach produces models that not only deliver highly accurate predictions but also offer clear insights into underlying physical processes. By integrating advanced techniques such as Bayesian neural networks and Monte Carlo dropout (Deshpande & Kuleshov, 2023; Bharadwaja et al., 2022; Galetzka et al., 2019), these endeavors significantly enhance model transparency, enabling more informed decision-making in safety-critical applications

6. Educational and collaborative platforms - including interdisciplinary workshops, joint research projects, and comprehensive cross-training initiatives - to train a new generation of researchers fluent in both ML and physical sciences, fostering long-term collaborative ecosystems. The integration of emerging technologies like PINNs, digital twins, surrogate modelling and Bayesian optimization, multi-physics and multi-scale simulation, adaptive and intelligent control systems offers immersive and tailored learning experiences (Joshi et al., 2024; Nyangon & Akintunde, 2024; Bukhari & Kumar, 2024; Tahkola et al., 2020).

These opportunities promote enduring collaborative ecosystems, bridging theoretical and practical gaps while facilitating the rapid dissemination of cutting-edge PIML methodologies across the electric machines and drives industry.

## 5. Conclusions

Recent advancements in PIML are revolutionizing the design and optimization of electrical machines and drives. This state-of-the-art review demonstrates that integrating physical laws directly into ML frameworks not only enhances modelling accuracy but also significantly reduces computational costs. The novel methodologies discussed, such as PINNs and hybrid data–physics approaches, offer robust solutions to the challenges of modelling complex electromagnetic phenomena and multi-scale interactions inherent in electrical machines. By embedding Maxwell's equations and other governing principles into the learning process, these methods achieve higher fidelity in simulations and more reliable performance predictions. Moreover, the shift towards mesh-free and domain decomposition techniques, along with the utilization of Bayesian optimization, highlights a clear trend towards creating adaptable and efficient design tools that can operate in real-time conditions. This convergence of physical modelling and advanced ML heralds a new era in electrical machine design that is both data efficient and physically consistent (Asef & Vagg, 2024; Nyangon, 2021b).

Building on the integration of physics with ML, recent research has successfully addressed key challenges such as data sparsity and the need for rapid design iteration. The developed algorithms have not only improved computational efficiency but have also expanded the range of applicable operating conditions by effectively capturing nonlinearity and hysteresis effects. For instance, the application of PINNs in capturing complex electromagnetic behavior has enabled the replacement of traditional, computationally intensive finite element methods in many cases. In addition, these approaches facilitate the seamless incorporation of multi-scale models, thereby providing comprehensive performance insights from local field variations to overall machine behavior. This integration also supports improved parameter estimation and fault diagnostics, which are critical for enhancing the reliability of electrical drives under dynamic operating conditions. Such advancements indicate that a unified framework combining both physics and data-driven learning can robustly address longstanding modelling challenges (Cheruku et al., 2025).

The ongoing developments in PIML represent a transformative shift in the field of electrical machine design and control. By leveraging physical constraints and domain knowledge, researchers are now able to develop models that are not only more accurate but also computationally efficient and scalable. The evidence from recent studies underscores that these methods can significantly reduce design time while achieving improvements in key performance metrics such as efficiency, torque, and noise reduction. Furthermore, the emerging synergy between physics-informed frameworks and optimization techniques promises to enhance the robustness of design processes across varied applications—from automotive drives to advanced industrial machinery. As future research continues to refine these methods and extend their applicability, it is anticipated that PIML will become a cornerstone technology for next-generation electrical machine design, ultimately contributing to more sustainable and high-performance energy systems (Nyangon, 2025a, 2021a).

Future research should concentrate on refining the integration of multi-physics constraints into neural architectures, thereby reducing training complexity and improving convergence. Furthermore, there is a need to develop adaptive methods that can dynamically adjust to real-time variations in system parameters, ultimately leading to more resilient predictive maintenance and control frameworks. Advancements in hardware acceleration and parallel computing will further enhance the capability of these methods, enabling real-time applications in industrial environments while ensuring rigorous adherence to physical laws. Finally, integrating data-driven techniques with traditional numerical simulations may further streamline the design optimization process, leading to cost-effective and scalable solutions.

## References

Abu-Rub, O. H., Fard, A. Y., Umar, M. F., Hosseinzadehtaher, M., & Shadmands, M. B. (2021). Towards Intelligent Power Electronics-Dominated Grid via Machine Learning Techniques. *IEEE Power Electronics Magazine*, *8*(1), 28–38. IEEE Power Electronics Magazine. https://doi.org/10.1109/MPEL.2020.3047506

Accetta, A., Cirrincione, M., Pucci, M., & Sferlazza, A. (2022). Feedback Linearization Based Nonlinear Control of SynRM Drives Accounting for Self- and Cross-Saturation. *IEEE Transactions on Industry Applications*, *58*(3), 3637–3651. IEEE Transactions on Industry Applications. https://doi.org/10.1109/TIA.2022.3155511

Achouch, M., Dimitrova, M., Ziane, K., Sattarpanah Karganroudi, S., Dhouib, R., Ibrahim, H., & Adda, M. (2022). On Predictive Maintenance in Industry 4.0: Overview, Models, and Challenges. *Applied Sciences*, *12*(16), Article 16. https://doi.org/10.3390/app12168081

Agamloh, E., von Jouanne, A., & Yokochi, A. (2020). An Overview of Electric Machine Trends in Modern Electric Vehicles. *Machines*, *8*(2), Article 2. https://doi.org/10.3390/machines8020020

Aissaoui, A. G. (2018). Introductory Chapter: Introduction to the Design and the Control of Electrical Machines. In *Optimization and Control of Electrical Machines*. IntechOpen. https://doi.org/10.5772/intechopen.78772

Al-Aghbari, M., & Gujarathi, A. M. (2022). Stochastic multi-objective optimization approaches in a real-world oil field waterflood management. *Journal of Petroleum Science and Engineering*, *218*, 110920. https://doi.org/10.1016/j.petrol.2022.110920

Alatawneh, N., & Pillay, P. (2016). Calibration of the Tangential Coil Sensor for the Measurement of Core Losses in Electrical Machine Laminations. *IEEE Transactions on Energy Conversion*, *31*(2), 413–423. IEEE Transactions on Energy Conversion. https://doi.org/10.1109/TEC.2016.2525011

Ali, M. H., Ramadan, A. A. R., Abass, D. M., & Ali, M. H. (2024). Analysis and simulation model of three phase Permanent Magnet Synchronous Motor Drive (PMSM). *E3S Web of Conferences*, *542*, 01011. https://doi.org/10.1051/e3sconf/202454201011

Aljafari, B., Ashok Kumar, L., Indragandhi, V., & Subramaniyaswamy, V. (2022). Analysis and Implementation of Sliding Mode Controller-Based Variable Frequency Drive Using the SCADA System. *International Transactions on Electrical Energy Systems*, *2022*(1), 7194119. https://doi.org/10.1155/2022/7194119

Antonelo, E. A., Camponogara, E., Seman, L. O., Jordanou, J. P., de Souza, E. R., & Hübner, J. F. (2024). Physics-informed neural nets for control of dynamical systems. *Neurocomputing*, *579*, 127419. https://doi.org/10.1016/j.neucom.2024.127419

Antonion, K., Wang, X., Raissi, M., & Joshie, L. (2024). Machine Learning Through Physics–Informed Neural Networks: Progress and Challenges. *Academic Journal of Science and Technology*, *9*(1), Article 1. https://doi.org/10.54097/b1d21816

Antunes, H. M. A., Brandao, D. I., Biajo, V. H. M., Alves, M. H. S., Oliveira, F. S., & Silva, S. M. (2023). Floating, Production, Storage, and Offloading Unit: A Case Study Using Variable Frequency Drives. *IEEE Transactions on Industry Applications*, *59*(4), 4764–4772. IEEE Transactions on Industry Applications. https://doi.org/10.1109/TIA.2023.3258420

Arzani, A., Wang, J.-X., & D’Souza, R. M. (2021). Uncovering near-wall blood flow from sparse data with physics-informed neural networks. *Physics of Fluids*, *33*(7), 071905. https://doi.org/10.1063/5.0055600

Asef, P., & Vagg, C. (2024). A physics-informed Bayesian optimization method for rapid development of electrical machines. *Scientific Reports*, *14*(1), 4526. https://doi.org/10.1038/s41598-024-54965-2

Bachchhav, P. P., Kasar, T. M., & Zete, R. S. (2017). Energy conservation by energy efficient drive. *2017 International Conference on Innovations in Information, Embedded and Communication Systems (ICIIECS)*, 1–6. https://doi.org/10.1109/ICIIECS.2017.8276053

Bahrami, M., & Khashroum, Z. (2023). Review of Machine Learning Techniques for Power Electronics Control and Optimization. *Computational Research Progress in Applied Science and Engineering*, *9*(3), 1–8. https://doi.org/10.61186/crpase.9.3.2860

Bajaj, C., McLennan, L., Andeen, T., & Roy, A. (2023). Recipes for when physics fails: Recovering robust learning of physics informed neural networks. *Machine Learning: Science and Technology*, *4*(1), 015013. https://doi.org/10.1088/2632-2153/acb416

Bharadwaja, B. V. S. S., Nabian, M. A., Sharma, B., Choudhry, S., & Alankar, A. (2022). Physics-Informed Machine Learning and Uncertainty Quantification for Mechanics of Heterogeneous Materials. *Integrating Materials and Manufacturing Innovation*, *11*(4), 607–627. https://doi.org/10.1007/s40192-022-00283-2

Bianchi, N., Babetto, C., & Bacco, G. (2021). *Synchronous reluctance motor optimization for pumping application*. https://digital-library.theiet.org/doi/abs/10.1049/PBPO186E_ch11

Biasion, M., Fernandes, J. F. P., Vaschetto, S., Cavagnino, A., & Tenconi, A. (2021). Superconductivity and its Application in the Field of Electrical Machines. *2021 IEEE International Electric Machines & Drives Conference (IEMDC)*, 1–7. https://doi.org/10.1109/IEMDC47953.2021.9449545

Biot-Monterde, V., Navarro-Navarro, Á., Antonino-Daviu, J. A., & Razik, H. (2021). Stray Flux Analysis for the Detection and Severity Categorization of Rotor Failures in Induction Machines Driven by Soft-Starters. *Energies*, *14*(18), Article 18. https://doi.org/10.3390/en14185757

Boldea, I., Tutelea, L. N., Xu, W., & Pucci, M. (2018). Linear Electric Machines, Drives, and MAGLEVs: An Overview. *IEEE Transactions on Industrial Electronics*, *65*(9), 7504–7515. IEEE Transactions on Industrial Electronics. https://doi.org/10.1109/TIE.2017.2733492

Book, G., Traue, A., Balakrishna, P., Brosch, A., Schenke, M., Hanke, S., Kirchgässner, W., & Wallscheid, O. (2021). Transferring Online Reinforcement Learning for Electric Motor Control From Simulation to Real-World Experiments. *IEEE Open Journal of Power Electronics*, *2*, 187–201. IEEE Open Journal of Power Electronics. https://doi.org/10.1109/OJPEL.2021.3065877

Bose, B. K. (2001). Artificial neural network applications in power electronics. *IECON'01. 27th Annual Conference of the IEEE Industrial Electronics Society (Cat. No.37243)*, *3*, 1631–1638 vol.3. https://doi.org/10.1109/IECON.2001.975533

Bose, B. K. (2005). Power Electronics and Motor Drives—Technology Advances, Trends and Applications. *2005 IEEE International Conference on Industrial Technology*, P20–P64. https://doi.org/10.1109/ICIT.2005.1600874

Bose, B. K. (2020). Artificial Intelligence Techniques: How Can it Solve Problems in Power Electronics?: An Advancing Frontier. *IEEE Power Electronics Magazine*, *7*(4), 19–27. IEEE Power Electronics Magazine. https://doi.org/10.1109/MPEL.2020.3033607

Bouzid, S., Viarouge, P., & Cros, J. (2020). Real-Time Digital Twin of a Wound Rotor Induction Machine Based on Finite Element Method. *Energies*, *13*(20), Article 20. https://doi.org/10.3390/en13205413

Brosch, A., Hanke, S., Wallscheid, O., & Böcker, J. (2021). Data-Driven Recursive Least Squares Estimation for Model Predictive Current Control of Permanent Magnet Synchronous Motors. *IEEE Transactions on Power Electronics*, *36*(2), 2179–2190. IEEE Transactions on Power Electronics. https://doi.org/10.1109/TPEL.2020.3006779

Bukhari, S. A. A. S., & Kumar, V. (2024). Development of a Synchronous Reluctance Motor for Industrial application. *Sukkur IBA Journal of Emerging Technologies*, *7*(1), Article 1. https://doi.org/10.30537/sjet.v7i1.1461

Carleo, G., Cirac, I., Cranmer, K., Daudet, L., Schuld, M., Tishby, N., Vogt-Maranto, L., & Zdeborová, L. (2019). Machine learning and the physical sciences. *Reviews of Modern Physics*, *91*(4), 045002. https://doi.org/10.1103/RevModPhys.91.045002

Cerqueira, V., Torgo, L., & Mozetič, I. (2020). Evaluating time series forecasting models: An empirical study on performance estimation methods. *Machine Learning*, *109*(11), 1997–2028. https://doi.org/10.1007/s10994-020-05910-7

Chen, J., Hu, W., Cao, D., Zhang, Z., Chen, Z., & Blaabjerg, F. (2023). A Meta-Learning Method for Electric Machine Bearing Fault Diagnosis Under Varying Working Conditions With Limited Data. *IEEE Transactions on Industrial Informatics*, *19*(3), 2552–2564. IEEE Transactions on Industrial Informatics. https://doi.org/10.1109/TII.2022.3165027

Chen, R., & Tong, T. (2023). Induction Motors and Permanent Magnet Motors in Electric Vehicles: Characteristics and Development Trends. *2023 International Conference on Internet of Things, Robotics and Distributed Computing (ICIRDC)*, 221–224. https://doi.org/10.1109/ICIRDC62824.2023.00046

Cheruku, S., Balaji, S., Delgado, A., & Krishnamurthy, V. R. (2025). Data-Driven Digital Twins for Real-Time Machine Monitoring: A Case Study on a Rotating Machine. *Journal of Computing and Information Science in Engineering*, *25*(031005). https://doi.org/10.1115/1.4067600

Çınar, Z. M., Abdussalam Nuhu, A., Zeeshan, Q., Korhan, O., Asmael, M., & Safaei, B. (2020). Machine Learning in Predictive Maintenance towards Sustainable Smart Manufacturing in Industry 4.0. *Sustainability*, *12*(19), Article 19. https://doi.org/10.3390/su12198211

Corral-Hernandez, J. A., & Antonino-Daviu, J. A. (2018). Thorough validation of a rotor fault diagnosis methodology in laboratory and field soft-started induction motors. *Chinese Journal of Electrical Engineering*, *4*(3), 66–72. Chinese Journal of Electrical Engineering. https://doi.org/10.23919/CJEE.2018.8471291

Cuomo, S., Di Cola, V. S., Giampaolo, F., Rozza, G., Raissi, M., & Piccialli, F. (2022). Scientific Machine Learning Through Physics–Informed Neural Networks: Where we are and What's Next. *Journal of Scientific Computing*, *92*(3), 88. https://doi.org/10.1007/s10915-022-01939-z

Dalzochio, J., Kunst, R., Pignaton, E., Binotto, A., Sanyal, S., Favilla, J., & Barbosa, J. (2020). Machine learning and reasoning for predictive maintenance in Industry 4.0: Current status and challenges. *Computers in Industry*, *123*, 103298. https://doi.org/10.1016/j.compind.2020.103298

De Doncker, R., Pulle, D. W. J., & Veltman, A. (2011). Modern Electrical Drives: An Overview. In R. De Doncker, D. W. J. Pulle, & A. Veltman (Eds.), *Advanced Electrical Drives: Analysis, Modeling, Control* (pp. 1–15). Springer Netherlands. https://doi.org/10.1007/978-94-007-0181-6_1

De Florio, M., Schiassi, E., & Furfaro, R. (2022). Physics-informed neural networks and functional interpolation for stiff chemical kinetics. *Chaos: An Interdisciplinary Journal of Nonlinear Science*, *32*(6), 063107. https://doi.org/10.1063/5.0086649

Deraz, S. A., & Azazi, H. Z. (2017). Current limiting soft starter for three phase induction motor drive system using PWM AC chopper. *IET Power Electronics*, *10*(11), 1298–1306. https://doi.org/10.1049/iet-pel.2016.0762

Deshpande, S., & Kuleshov, V. (2023). *Calibrated Uncertainty Estimation Improves Bayesian Optimization* (No. arXiv:2112.04620; Version 3). arXiv. https://doi.org/10.48550/arXiv.2112.04620

Diao, K., Sun, X., Bramerdorfer, G., Cai, Y., Lei, G., & Chen, L. (2022). Design optimization of switched reluctance machines for performance and reliability enhancements: A review. *Renewable and Sustainable Energy Reviews*, *168*, 112785. https://doi.org/10.1016/j.rser.2022.112785

Diz, S. de L., López, R. M., Sánchez, F. J. R., Llerena, E. D., & Peña, E. J. B. (2023). A real-time digital twin approach on three-phase power converters applied to condition monitoring. *Applied Energy*, *334*, 120606. https://doi.org/10.1016/j.apenergy.2022.120606

Donaghy-Spargo, C. (2016). Synchronous reluctance motor technology: Industrial opportunities, challenges and future direction. *Engineering & Technology Reference*, *2016*. https://doi.org/10.1049/etr.2015.0044

Elmorshedy, M. F., Xu, W., El-Sousy, F. F. M., Islam, Md. R., & Ahmed, A. A. (2021). Recent Achievements in Model Predictive Control Techniques for Industrial Motor: A Comprehensive State-of-the-Art. *IEEE Access*, *9*, 58170–58191. IEEE Access. https://doi.org/10.1109/ACCESS.2021.3073020

Elsaid, K., Safar, M., & El-Kharashi, M. W. (2022). Optimized FPGA Architecture for Machine Learning Applications using Posit Multipliers. *2022 International Conference on Microelectronics (ICM)*, 50–53. https://doi.org/10.1109/ICM56065.2022.10005431

Escapil-Inchauspé, P., & Ruz, G. A. (2023). H-Analysis and data-parallel physics-informed neural networks. *Scientific Reports*, *13*(1), 17562. https://doi.org/10.1038/s41598-023-44541-5

Faiz, J., Ghorbanian, V., & Joksimovic, G. (2017). *Fault Diagnosis of Induction Motors*. https://doi.org/10.1049/PBPO108E

Falekas, G., & Karlis, A. (2021). Digital Twin in Electrical Machine Control and Predictive Maintenance: State-of-the-Art and Future Prospects. *Energies*, *14*(18), Article 18. https://doi.org/10.3390/en14185933

Farah, N., Lei, G., Zhu, J., & Guo, Y. (2023). Reinforcement Learning for Intelligent Control of AC Machine Drives: A Review. *2023 IEEE International Future Energy Electronics Conference (IFEEC)*, 1–6. https://doi.org/10.1109/IFEEC58486.2023.10458623

Feng, Z., Liang, W., Ling, J., Xiao, X., Tan, K. K., & Lee, T. H. (2020). Integral terminal sliding-mode-based adaptive integral backstepping control for precision motion of a piezoelectric ultrasonic motor. *Mechanical Systems and Signal Processing*, *144*, 106856. https://doi.org/10.1016/j.ymssp.2020.106856

Flieh, H., Lorenz, R. D., Totoki, E., Yamaguchi, S., & Nakamura, Y. (2017). Investigation of different servo motor designs for servo cycle operations and loss minimizing control performance. *2017 IEEE Energy Conversion Congress and Exposition (ECCE)*, 4316–4323. https://doi.org/10.1109/ECCE.2017.8096744

Foo, G. H. B., & Zhang, X. (2016). Constant Switching Frequency Based Direct Torque Control of Interior Permanent Magnet Synchronous Motors With Reduced Ripples and Fast Torque Dynamics. *IEEE Transactions on Power Electronics*, *31*(9), 6485–6493. IEEE Transactions on Power Electronics. https://doi.org/10.1109/TPEL.2015.2503292

Fukushige, T., Limsuwan, N., Kato, T., Akatsu, K., & Lorenz, R. D. (2013). Efficiency contours and loss minimization over a driving cycle of a variable-flux flux-intensifying interior permanent magnet machine. *2013 IEEE Energy Conversion Congress and Exposition*, 591–597. https://doi.org/10.1109/ECCE.2013.6646755

Gaber, M., El-Banna, S. H., El-Dabah, M., & Hamad, M. S. (2021). Intelligent Energy Management System for an all-electric ship based on adaptive neuro-fuzzy inference system. *Energy Reports*, *7*, 7989–7998. https://doi.org/10.1016/j.egyr.2021.06.054

Galetzka, A., Bontinck, Z., Römer, U., & Schöps, S. (2019). A Multilevel Monte Carlo Method for High-Dimensional Uncertainty Quantification of Low-Frequency Electromagnetic Devices. *IEEE Transactions on Magnetics*, *55*(8), 1–12. IEEE Transactions on Magnetics. https://doi.org/10.1109/TMAG.2019.2911053

Gawande, R. M. (2024). Machine Learning Approaches for Fault Detection and Diagnosis in Electrical Machines: A Comparative Study of Deep Learning and Classical Methods. *Panamerican Mathematical Journal*, *34*(2), Article 2. https://doi.org/10.52783/pmj.v34.i2.930

Ghattas, O., & Willcox, K. (2021). Learning physics-based models from data: Perspectives from inverse problems and model reduction. *Acta Numerica*, *30*, 445–554. https://doi.org/10.1017/S0962492921000064

Green, S., Vineyard, C. M., & Koç, Ç. K. (2018). Mathematical Optimizations for Deep Learning. In Ç. K. Koç (Ed.), *Cyber-Physical Systems Security* (pp. 69–92). Springer International Publishing. https://doi.org/10.1007/978-3-319-98935-8_4

Gu, L., Qin, S., Xu, L., & Chen, R. (2024). Physics-informed neural networks with domain decomposition for the incompressible Navier–Stokes equations. *Physics of Fluids*, *36*(2), 021914. https://doi.org/10.1063/5.0188830

Guo, J., Quéval, L., Roucaries, B., Vido, L., Liu, L., Trillaud, F., & Berriaud, C. (2020). Nonlinear Current Sheet Model of Electrical Machines. *IEEE Transactions on Magnetics*, *56*(1), 1–4. IEEE Transactions on Magnetics. https://doi.org/10.1109/TMAG.2019.2950614

Guo, X., Hu, X., & Zhang, S. (2024). Application status of variable-frequency drive in hydrogen fuel cell air compressors from an industrial viewpoint: A review. *Sustainable Energy Technologies and Assessments*, *64*, 103716. https://doi.org/10.1016/j.seta.2024.103716

Gupta, G., Sudeep, R., Ashok, B., Vignesh, R., Kannan, C., Kavitha, C., Alroobaea, R., Alsafyani, M., AboRas, K. M., & Emara, A. (2024). Intelligent Regenerative Braking Control With Novel Friction Coefficient Estimation Strategy for Improving the Performance Characteristics of Hybrid Electric Vehicle. *IEEE Access*, *12*, 110361–110384. IEEE Access. https://doi.org/10.1109/ACCESS.2024.3440210

Gupta, S., Jain, D. S., Roy, B., & Deb, A. (2022). A TinyML Approach to Human Activity Recognition. *Journal of Physics: Conference Series*, *2273*(1), 012025. https://doi.org/10.1088/1742-6596/2273/1/012025

Hakami, S. S., Alsofyani, I. M., & Lee, K.-B. (2019). Torque Ripple Reduction and Flux-Droop Minimization of DTC With Improved Interleaving CSFTC of IM Fed by Three-Level NPC Inverter. *IEEE Access*, *7*, 184266–184275. IEEE Access. https://doi.org/10.1109/ACCESS.2019.2960685

Hamolia, V., & Melnyk, V. (2021). A Survey of Machine Learning Methods and Applications in Electronic Design Automation. *2021 11th International Conference on Advanced Computer Information Technologies (ACIT)*, 757–760. https://doi.org/10.1109/ACIT52158.2021.9548117

Han, S., Mao, H., & Dally, W. (2015). Deep Compression: Compressing Deep Neural Network with Pruning, Trained Quantization and Huffman Coding. *arXiv: Computer Vision and*

*Pattern Recognition*. https://www.semanticscholar.org/paper/642d0f49b7826adcf986616f4af77e736229990f

Hanover, D., Foehn, P., Sun, S., Kaufmann, E., & Scaramuzza, D. (2022). Performance, Precision, and Payloads: Adaptive Nonlinear MPC for Quadrotors. *IEEE Robotics and Automation Letters*, *7*(2), 690–697. IEEE Robotics and Automation Letters. https://doi.org/10.1109/LRA.2021.3131690

Hao, Z., Liu, S., Zhang, Y., Ying, C., Feng, Y., Su, H., & Zhu, J. (2023). *Physics-Informed Machine Learning: A Survey on Problems, Methods and Applications* (No. arXiv:2211.08064). arXiv. https://doi.org/10.48550/arXiv.2211.08064

Hawks, B., Duarte, J., Fraser, N. J., Pappalardo, A., Tran, N., & Umuroglu, Y. (2021). Ps and Qs: Quantization-Aware Pruning for Efficient Low Latency Neural Network Inference. *Frontiers in Artificial Intelligence*, *4*. https://doi.org/10.3389/frai.2021.676564

Hossein Rahimighazvini, Zeyad Khashroum, Maryam Bahrami, & Milad Hadizadeh Masali. (2024). Power electronics anomaly detection and diagnosis with machine learning and deep learning methods: A survey. *International Journal of Science and Research Archive*, *11*(2), 730–739. https://doi.org/10.30574/ijsra.2024.11.2.0428

Hsieh, H.-L., Wong, Y. T., Pesaran, B., & Shanechi, M. M. (2018). Multiscale modeling and decoding algorithms for spike-field activity. *Journal of Neural Engineering*, *16*(1), 016018. https://doi.org/10.1088/1741-2552/aaeb1a

Huber, M. C., Fuhrländer, M., & Schöps, S. (2023). Multi-Objective Yield Optimization for Electrical Machines Using Gaussian Processes to Learn Faulty Design. *IEEE Transactions on Industry Applications*, *59*(2), 1340–1350. IEEE Transactions on Industry Applications. https://doi.org/10.1109/TIA.2022.3211250

Husebø, A. B., Khang, H. V., & Pawlus, W. (2019). Diagnosis of Incipient Bearing Faults using Convolutional Neural Networks. *2019 IEEE Workshop on Electrical Machines Design, Control and Diagnosis (WEMDCD)*, *1*, 143–149. https://doi.org/10.1109/WEMDCD.2019.8887785

Itagi, A., Krishvadana, S., Bharath, K. P., & Rajesh Kumar, M. (2021). FPGA Architecture To Enhance Hardware Acceleration for Machine Learning Applications. *2021 5th International Conference on Computing Methodologies and Communication (ICCMC)*, 1716–1722. https://doi.org/10.1109/ICCMC51019.2021.9418015

Jagtap, A. D., & Karniadakis, G. E. (2020). Extended Physics-Informed Neural Networks (XPINNs): A Generalized Space-Time Domain Decomposition Based Deep Learning Framework for Nonlinear Partial Differential Equations. *Communications in Computational Physics*, *31*(5), 2002–2041. https://doi.org/10.4208/cicp.OA-2020-0164

Jansson, E., Thiringer, T., & Grunditz, E. (2020). Convergence of Core Losses in a Permanent Magnet Machine, as Function of Mesh Density Distribution, a Case-Study Using Finite-Element Analysis. *IEEE Transactions on Energy Conversion*, *35*(3), 1667–1675. IEEE Transactions on Energy Conversion. https://doi.org/10.1109/TEC.2020.2982265

Ji, X., & Abdoli, S. (2023). Challenges and Opportunities in Product Life Cycle Management in the Context of Industry 4.0. *Procedia CIRP*, *119*, 29–34. https://doi.org/10.1016/j.procir.2023.04.002

Joshi, K., Snigdha, V., & Bhattacharya, A. K. (2024). Physics Informed Extreme Learning Machines With Residual Variation Diminishing Scheme for Nonlinear Problems With Discontinuous Surfaces. *IEEE Access*, *12*, 130617–130629. IEEE Access. https://doi.org/10.1109/ACCESS.2024.3457670

Kapp, S., Wang, C., McNelly, M., Romeiko, X., & Choi, J.-K. (2024). A comprehensive analysis of the energy, economic, and environmental impacts of industrial variable frequency drives. *Journal of Cleaner Production*, *434*, 140474. https://doi.org/10.1016/j.jclepro.2023.140474

Kargah-Ostadi, N., Vasylevskyi, K., Ablets, A., & Drach, A. (2024). Physics-informed neural networks to advance pavement engineering and management. *Road Materials and Pavement Design*, *25*(11), 2382–2403. https://doi.org/10.1080/14680629.2024.2315073

Karniadakis, G. E., Kevrekidis, I. G., Lu, L., Perdikaris, P., Wang, S., & Yang, L. (2021). Physics-informed machine learning. *Nature Reviews Physics*, *3*(6), 422–440. https://doi.org/10.1038/s42254-021-00314-5

Kathail, V. (2020). Xilinx Vitis Unified Software Platform. *Proceedings of the 2020 ACM/SIGDA International Symposium on Field-Programmable Gate Arrays*, 173–174. https://doi.org/10.1145/3373087.3375887

Kazemikia, D. (2024). *Reinforcement Learning for Motor Control: A Comprehensive Review* (No. arXiv:2412.17936). arXiv. https://doi.org/10.48550/arXiv.2412.17936

Khademi, A., & Dufour, S. (2024). A novel discretized physics-informed neural network model applied to the Navier–Stokes equations. *Physica Scripta*, *99*(7), 076016. https://doi.org/10.1088/1402-4896/ad5592

Khashroum, Z., Rahimighazvini, H., & Bahrami, M. (2023). Applications of Machine Learning in Power Electronics: A Specialization on Convolutional Neural Networks. *ENG Transactions*, *4*(1), 1–5. https://doi.org/10.61186/engt.4.1.2866

Kirchgässner, W., Wallscheid, O., & Böcker, J. (2021). Data-Driven Permanent Magnet Temperature Estimation in Synchronous Motors With Supervised Machine Learning: A Benchmark. *IEEE Transactions on Energy Conversion*, *36*(3), 2059–2067. IEEE Transactions on Energy Conversion. https://doi.org/10.1109/TEC.2021.3052546

Kitayoshi, R., Yoshiura, Y., & Kaku, Y. (2023). Σ-X Series: AC Servo Drive for Achievement of Digital Solution. *IEEJ Journal of Industry Applications*, *12*(5), 859–867. https://doi.org/10.1541/ieejjia.22010390

Kovacs, A., Fischbacher, J., Gusenbauer, M., Oezelt, H., Herper, H. C., Vekilova, O. Yu., Nieves, P., Arapan, S., & Schrefl, T. (2020). Computational Design of Rare-Earth Reduced Permanent Magnets. *Engineering*, *6*(2), 148–153. https://doi.org/10.1016/j.eng.2019.11.006

Koyama, S., Ribeiro, J. G. C., Nakamura, T., Ueno, N., & Pezzoli, M. (2024). Physics-Informed Machine Learning for Sound Field Estimation: Fundamentals, state of the art, and challenges. *IEEE Signal Processing Magazine*, *41*(6), 60–71. IEEE Signal Processing Magazine. https://doi.org/10.1109/MSP.2024.3465896

Kudelina, K., Vaimann, T., Asad, B., Rassõlkin, A., Kallaste, A., & Demidova, G. (2021). Trends and Challenges in Intelligent Condition Monitoring of Electrical Machines Using Machine Learning. *Applied Sciences*, *11*(6), Article 6. https://doi.org/10.3390/app11062761

Kumar, P., & Shankar Hati, A. (2021). Convolutional neural network with batch normalisation for fault detection in squirrel cage induction motor. *IET Electric Power Applications*, *15*(1), 39–50. https://doi.org/10.1049/elp2.12005

Kustowski, B., Gaffney, J. A., Spears, B. K., Anderson, G. J., Anirudh, R., Bremer, P.-T., Thiagarajan, J. J., G Kruse, M. K., & Nora, R. C. (2022). Suppressing simulation bias in multi-modal data using transfer learning. *Machine Learning: Science and Technology*, *3*(1), 015035. https://doi.org/10.1088/2632-2153/ac5e3e

Laghi, L., Schiassi, E., De Florio, M., Furfaro, R., & Mostacci, D. (2023). Physics-Informed Neural Networks for 1-D Steady-State Diffusion-Advection-Reaction Equations. *Nuclear*

*Science and Engineering*, *197*(9), 2373–2403. https://doi.org/10.1080/00295639.2022.2160604

Lambard, G., Sasaki, T. T., Sodeyama, K., Ohkubo, T., & Hono, K. (2022). Optimization of direct extrusion process for Nd-Fe-B magnets using active learning assisted by machine learning and Bayesian optimization. *Scripta Materialia*, *209*, 114341. https://doi.org/10.1016/j.scriptamat.2021.114341

Leake, C., & Mortari, D. (2020). Deep Theory of Functional Connections: A New Method for Estimating the Solutions of Partial Differential Equations. *Machine Learning and Knowledge Extraction*, *2*(1), Article 1. https://doi.org/10.3390/make2010004

Lei, G., Zhu, J., Guo, Y., Liu, C., & Ma, B. (2017). A Review of Design Optimization Methods for Electrical Machines. *Energies*, *10*(12), Article 12. https://doi.org/10.3390/en10121962

Li, D., Kakosimos, P., & Peretti, L. (2023). Machine-Learning-Based Condition Monitoring of Power Electronics Modules in Modern Electric Drives. *IEEE Power Electronics Magazine*, *10*(1), 58–66. IEEE Power Electronics Magazine. https://doi.org/10.1109/MPEL.2023.3236462

Li, S., Liu, C., & Ni, H. (2024). Enhancing neural operator learning with invariants to simultaneously learn various physical mechanisms. *National Science Review*, *11*(8), nwae198. https://doi.org/10.1093/nsr/nwae198

Li, S., Zhang, S., Habetler, T. G., & Harley, R. G. (2019). Modeling, Design Optimization, and Applications of Switched Reluctance Machines—A Review. *IEEE Transactions on Industry Applications*, *55*(3), 2660–2681. IEEE Transactions on Industry Applications. https://doi.org/10.1109/TIA.2019.2897965

Linka, K., Schäfer, A., Meng, X., Zou, Z., Karniadakis, G. E., & Kuhl, E. (2022). Bayesian Physics Informed Neural Networks for real-world nonlinear dynamical systems. *Computer Methods in Applied Mechanics and Engineering*, *402*, 115346. https://doi.org/10.1016/j.cma.2022.115346

Liu, D., & Wang, Y. (2019). Multi-Fidelity Physics-Constrained Neural Network and Its Application in Materials Modeling. *Journal of Mechanical Design*, *141*(121403). https://doi.org/10.1115/1.4044400

Liu, L., Guo, Y., Yin, W., Lei, G., & Zhu, J. (2022). Design and Optimization Technologies of Permanent Magnet Machines and Drive Systems Based on Digital Twin Model. *Energies*, *15*(17), Article 17. https://doi.org/10.3390/en15176186

Liu, W., & Pyrcz, M. J. (2023). Physics-informed graph neural network for spatial-temporal production forecasting. *Geoenergy Science and Engineering*, *223*, 211486. https://doi.org/10.1016/j.geoen.2023.211486

Loureiro, R. B., Sá, P. H. M., Lisboa, F. V. N., Peixoto, R. M., Nascimento, L. F. S., Bonfim, Y. da S., Cruz, G. O. R., Ramos, T. de O., Montes, C. H. R. L., Pagano, T. P., Pinheiro, O. R., & Borges, R. (2023). Efficient Deployment of Machine Learning Models on Microcontrollers: A Comparative Study of Quantization And Pruning Strategies. *Blucher Engineering Proceedings*, *10*(5), 181–188. https://doi.org/10.5151/siintec2023-305873

Ma, X., Zhang, Z., & Hua, H. (2022). Uncertainty quantization and reliability analysis for rotor/stator rub-impact using advanced Kriging surrogate model. *Journal of Sound and Vibration*, *525*, 116800. https://doi.org/10.1016/j.jsv.2022.116800

Mackay, C. T., & Nowell, D. (2023). Informed machine learning methods for application in engineering: A review. *Proceedings of the Institution of Mechanical Engineers, Part C: Journal of Mechanical Engineering Science*, *237*(24), 5801–5818. https://doi.org/10.1177/09544062231164575

Manfredi, P., & Trinchero, R. (2022). A Probabilistic Machine Learning Approach for the Uncertainty Quantification of Electronic Circuits Based on Gaussian Process Regression. *IEEE Transactions on Computer-Aided Design of Integrated Circuits and Systems*, *41*(8), 2638–2651. IEEE Transactions on Computer-Aided Design of Integrated Circuits and Systems. https://doi.org/10.1109/TCAD.2021.3112138

Mao, C., & Jin, Y. (2024). Uncertainty quantification study of the physics-informed machine learning models for critical heat flux prediction. *Progress in Nuclear Energy*, *170*, 105097. https://doi.org/10.1016/j.pnucene.2024.105097

Marculescu, D., Stamoulis, D., & Cai, E. (2018). Hardware-Aware Machine Learning: Modeling and Optimization. *2018 IEEE/ACM International Conference on Computer-Aided Design (ICCAD)*, 1–8. https://doi.org/10.1145/3240765.3243479

Marian, M., & Tremmel, S. (2023). Physics-Informed Machine Learning—An Emerging Trend in Tribology. *Lubricants*, *11*(11), Article 11. https://doi.org/10.3390/lubricants11110463

Mayr, A., Meyer, A., Seefried, J., Weigelt, M., Lutz, B., Sultani, D., Hampl, M., & Franke, J. (2017). Potentials of machine learning in electric drives production using the example of contacting processes and selective magnet assembly. *2017 7th International Electric Drives Production Conference (EDPC)*, 1–8. https://doi.org/10.1109/EDPC.2017.8328166

Mbula, B. L., & Chowdhury, D. (2017). Performance improvement of synchronous reluctance motors: A review. *2017 IEEE PES PowerAfrica*, 402–406. https://doi.org/10.1109/PowerAfrica.2017.7991258

Meesala, K. R. E., & Thippiripati, V. K. (2020). An Improved Direct Torque Control of Three-Level Dual Inverter Fed Open-Ended Winding Induction Motor Drive Based on Modified Look-Up Table. *IEEE Transactions on Power Electronics*, *35*(4), 3906–3917. IEEE Transactions on Power Electronics. https://doi.org/10.1109/TPEL.2019.2937684

Meng, C., Seo, S., Cao, D., Griesemer, S., & Liu, Y. (2022). *When Physics Meets Machine Learning: A Survey of Physics-Informed Machine Learning* (No. arXiv:2203.16797). arXiv. https://doi.org/10.48550/arXiv.2203.16797

Meriem, H., Nora, H., & Samir, O. (2023). Predictive Maintenance for Smart Industrial Systems: A Roadmap. *Procedia Computer Science*, *220*, 645–650. https://doi.org/10.1016/j.procs.2023.03.082

Milton, M., O, C. D. L., Ginn, H. L., & Benigni, A. (2020). Controller-Embeddable Probabilistic Real-Time Digital Twins for Power Electronic Converter Diagnostics. *IEEE Transactions on Power Electronics*, *35*(9), 9850–9864. IEEE Transactions on Power Electronics. https://doi.org/10.1109/TPEL.2020.2971775

Mistry, R., Mistry, B., & Lawrence, W. G. (2024). Why are Class I/Division 1 and Zone 1 Induction Machines Employed less and less in Industries Today? *2024 IEEE IAS Petroleum and Chemical Industry Technical Conference (PCIC)*, 1–7. https://doi.org/10.1109/PCIC47799.2024.10832201

Mohanraj, D., Aruldavid, R., Verma, R., Sathiyasekar, K., Barnawi, A. B., Chokkalingam, B., & Mihet-Popa, L. (2022). A Review of BLDC Motor: State of Art, Advanced Control Techniques, and Applications. *IEEE Access*, *10*, 54833–54869. IEEE Access. https://doi.org/10.1109/ACCESS.2022.3175011

Mojlish, S., Erdogan, N., Levine, D., & Davoudi, A. (2017). Review of Hardware Platforms for Real-Time Simulation of Electric Machines. *IEEE Transactions on Transportation Electrification*, *3*(1), 130–146. IEEE Transactions on Transportation Electrification. https://doi.org/10.1109/TTE.2017.2656141

Molnar, J. P., & Grauer, S. J. (2022). Flow field tomography with uncertainty quantification using a Bayesian physics-informed neural network. *Measurement Science and Technology*, *33*(6), 065305. https://doi.org/10.1088/1361-6501/ac5437

Moradi, S., Duran, B., Eftekhar Azam, S., & Mofid, M. (2023). Novel Physics-Informed Artificial Neural Network Architectures for System and Input Identification of Structural Dynamics PDEs. *Buildings*, *13*(3), Article 3. https://doi.org/10.3390/buildings13030650

Moseley, B., Markham, A., & Nissen-Meyer, T. (2023). Finite basis physics-informed neural networks (FBPINNs): A scalable domain decomposition approach for solving differential equations. *Advances in Computational Mathematics*, *49*(4), 62. https://doi.org/10.1007/s10444-023-10065-9

Murataliyev, M., Degano, M., Di Nardo, M., Bianchi, N., & Gerada, C. (2022). Synchronous Reluctance Machines: A Comprehensive Review and Technology Comparison. *Proceedings of the IEEE*, *110*(3), 382–399. Proceedings of the IEEE. https://doi.org/10.1109/JPROC.2022.3145662

Murphey, Y. L., Masrur, M. A., & Chen, Z. (2006). Fault Diagnostics in Electric Drives Using Machine Learning. In M. Ali & R. Dapoigny (Eds.), *Advances in Applied Artificial Intelligence* (pp. 1169–1178). Springer. https://doi.org/10.1007/11779568_124

Nagao, M., Datta-Gupta, A., Onishi, T., & Sankaran, S. (2024). Physics Informed Machine Learning for Reservoir Connectivity Identification and Robust Production Forecasting. *SPE Journal*, *29*(09), 4527–4541. https://doi.org/10.2118/219773-PA

Nan, F., & Hutter, M. (2024). Learning Adaptive Controller for Hydraulic Machinery Automation. *IEEE Robotics and Automation Letters*, *9*(4), 3972–3979. IEEE Robotics and Automation Letters. https://doi.org/10.1109/LRA.2024.3372831

Nannen, H., Zatocil, H., & Griepentrog, G. (2022). Predictive Firing Algorithm for Soft Starter Driven Induction Motors. *IEEE Transactions on Industrial Electronics*, *69*(12), 12152–12161. IEEE Transactions on Industrial Electronics. https://doi.org/10.1109/TIE.2021.3135606

Negri, E., Berardi, S., Fumagalli, L., & Macchi, M. (2020). MES-integrated digital twin frameworks. *Journal of Manufacturing Systems*, *56*, 58–71. https://doi.org/10.1016/j.jmsy.2020.05.007

Nghiem, T. X., Drgoňa, J., Jones, C., Nagy, Z., Schwan, R., Dey, B., Chakrabarty, A., Di Cairano, S., Paulson, J. A., Carron, A., Zeilinger, M. N., Shaw Cortez, W., & Vrabie, D. L. (2023). Physics-Informed Machine Learning for Modeling and Control of Dynamical Systems. *2023 American Control Conference (ACC)*, 3735–3750. https://doi.org/10.23919/ACC55779.2023.10155901

Nyangon, J. (2021a). Smart energy frameworks for smart cities: The need for polycentrism. In J. C. Augusto (Ed.), *Handbook of Smart Cities* (Vol. 1–Book, Section, pp. 1–32). Springer. https://doi.org/10.1007/978-3-030-15145-4_4-2

Nyangon, J. (2021b). Tackling the risk of stranded electricity assets with machine learning and Artificial Intelligence. In J. Nyangon & J. Byrne (Eds.), *Sustainable Energy Investment: Technical, Market and Policy Innovations to Address Risk* (Vol. 1–Book, Section, pp. 1–22). IntechOpen.

Nyangon, J. (2024). Climate-Proofing Critical Energy Infrastructure: Smart Grids, Artifici al Intelligence, and Machine Learning for Power System Resilience agai nst Extreme Weather Events. *Journal of Infrastructure Systems*, *30*(1). Crossref. https://doi.org/10.1061/jitse4.iseng-2375

Nyangon, J. (2025a). Physics informed neural networks for maritime energy systems and blue economy innovations. *Machine Learning: Earth*, *1*(1), 011002. https://doi.org/10.1088/3049-4753/adfe73
Nyangon, J. (2025b). Smart Grid Strategies for Tackling the Duck Curve: A Qualitative Assessment of Digitalization, Battery Energy Storage, and Managed Rebound Effects Benefits. *Energies*, *18*(15), 3988. https://doi.org/10.3390/en18153988
Nyangon, J., & Akintunde, R. (2024). Principal component analysis of day-ahead electricity price forecasting in CAISO and its implications for highly integrated renewable energy markets. *WIREs Energy and Environment*, e504. https://doi.org/10.1002/wene.504
Nyangon, J., & Darekar, A. (2024). Advancements in hydrogen energy systems: A review of levelized costs, financial incentives and technological innovations. *Innovation and Green Development*, *3*(3), 100149. https://doi.org/10.1016/j.igd.2024.100149
Öksüztepe, E., Kaya, U., & Kurum, H. (2022). A review of conventional and new-generation aircraft starter/generators in perspective of electric drive applications. *Aircraft Engineering and Aerospace Technology*, *95*(3), 474–487. https://doi.org/10.1108/AEAT-05-2022-0126
Osta, M., Alameh, M., Younes, H., Ibrahim, A., & Valle, M. (2019). Energy Efficient Implementation of Machine Learning Algorithms on Hardware Platforms. *2019 26th IEEE International Conference on Electronics, Circuits and Systems (ICECS)*, 21–24. https://doi.org/10.1109/ICECS46596.2019.8965157
Ozcelik, N. G., Dogru, U. E., Imeryuz, M., & Ergene, L. T. (2019). Synchronous Reluctance Motor vs. Induction Motor at Low-Power Industrial Applications: Design and Comparison. *Energies*, *12*(11), Article 11. https://doi.org/10.3390/en12112190
Parekh, V., Flore, D., & Schöps, S. (2023). Performance Analysis of Electrical Machines Using a Hybrid Data- and Physics-Driven Model. *IEEE Transactions on Energy Conversion*, *38*(1), 530–539. IEEE Transactions on Energy Conversion. https://doi.org/10.1109/TEC.2022.3209103
Parra, A., Zubizarreta, A., Pérez, J., & Dendaluce, M. (2018). Intelligent Torque Vectoring Approach for Electric Vehicles with Per-Wheel Motors. *Complexity*, *2018*(1), 7030184. https://doi.org/10.1155/2018/7030184
Pateras, J., Rana, P., & Ghosh, P. (2023). A Taxonomic Survey of Physics-Informed Machine Learning. *Applied Sciences*, *13*(12), Article 12. https://doi.org/10.3390/app13126892
Pereida, K., Kooijman, D., Duivenvoorden, R. R. P. R., & Schoellig, A. P. (2019). Transfer learning for high-precision trajectory tracking through adaptive feedback and iterative learning. *International Journal of Adaptive Control and Signal Processing*, *33*(2), 388–409. https://doi.org/10.1002/acs.2887
Prabhu, N., Thirumalaivasan, R., & Ashok, B. (2023). Critical Review on Torque Ripple Sources and Mitigation Control Strategies of BLDC Motors in Electric Vehicle Applications. *IEEE Access*, *11*, 115699–115739. IEEE Access. https://doi.org/10.1109/ACCESS.2023.3324419
Praslicka, B., Ma, C., & Taran, N. (2023). A Computationally Efficient High-Fidelity Multi-Physics Design Optimization of Traction Motors for Drive Cycle Loss Minimization. *IEEE Transactions on Industry Applications*, *59*(2), 1351–1360. IEEE Transactions on Industry Applications. https://doi.org/10.1109/TIA.2022.3220554
Rabbi, S. F., Kahnamouei, J. T., Liang, X., & Yang, J. (2020). Shaft Failure Analysis in Soft-Starter Fed Electrical Submersible Pump Systems. *IEEE Open Journal of Industry Applications*, *1*, 1–10. IEEE Open Journal of Industry Applications. https://doi.org/10.1109/OJIA.2019.2956987

Raia, M. R., Ciceo, S., Chauvicourt, F., & Martis, C. (2023). Multi-Attribute Machine Learning Model for Electrical Motors Performance Prediction. *Applied Sciences*, *13*(3), Article 3. https://doi.org/10.3390/app13031395

Raissi, M., Perdikaris, P., & Karniadakis, G. E. (2019). Physics-informed neural networks: A deep learning framework for solving forward and inverse problems involving nonlinear partial differential equations. *Journal of Computational Physics*, *378*, 686–707.

Rajasekar, V., & Ashok, B. (2024). Improvised energy management control through neuro-fuzzy based adaptive ECMS approach for an optimal battery utilization in non-plugin parallel hybrid electric vehicle. *Proceedings of the Institution of Mechanical Engineers, Part C: Journal of Mechanical Engineering Science*, *238*(8), 3308–3326. https://doi.org/10.1177/09544062231207186

Reagen, B., Adolf, R., Whatmough, P., Wei, G.-Y., & Brooks, D. (2017). Foundations of Deep Learning. In B. Reagen, R. Adolf, P. Whatmough, G.-Y. Wei, & D. Brooks (Eds.), *Deep Learning for Computer Architects* (pp. 9–23). Springer International Publishing. https://doi.org/10.1007/978-3-031-01756-8_2

Rohrhofer, F. M., Posch, S., Gößnitzer, C., & Geiger, B. C. (2023). Data vs. Physics: The Apparent Pareto Front of Physics-Informed Neural Networks. *IEEE Access*, *11*, 86252–86261. IEEE Access. https://doi.org/10.1109/ACCESS.2023.3302892

Rosofsky, S. G., Al Majed, H., & Huerta, E. A. (2023). Applications of physics informed neural operators. *Machine Learning: Science and Technology*, *4*(2), 025022. https://doi.org/10.1088/2632-2153/acd168

Roubache, L., Boughrara, K., Dubas, F., & Ibtiouen, R. (2018). New Subdomain Technique for Electromagnetic Performances Calculation in Radial-Flux Electrical Machines Considering Finite Soft-Magnetic Material Permeability. *IEEE Transactions on Magnetics*, *54*(4), 1–15. IEEE Transactions on Magnetics. https://doi.org/10.1109/TMAG.2017.2785254

Ruderman, M., Iwasaki, M., & Chen, W.-H. (2020). Motion-control techniques of today and tomorrow: A review and discussion of the challenges of controlled motion. *IEEE Industrial Electronics Magazine*, *14*(1), 41–55. IEEE Industrial Electronics Magazine. https://doi.org/10.1109/MIE.2019.2956683

Schenke, M., Kirchgässner, W., & Wallscheid, O. (2020). Controller Design for Electrical Drives by Deep Reinforcement Learning: A Proof of Concept. *IEEE Transactions on Industrial Informatics*, *16*(7), 4650–4658. IEEE Transactions on Industrial Informatics. https://doi.org/10.1109/TII.2019.2948387

Schiassi, E., De Florio, M., D'Ambrosio, A., Mortari, D., & Furfaro, R. (2021). Physics-Informed Neural Networks and Functional Interpolation for Data-Driven Parameters Discovery of Epidemiological Compartmental Models. *Mathematics*, *9*(17), Article 17. https://doi.org/10.3390/math9172069

Schüller, S., Yilmaz, S., & De Doncker, R. W. (2024). Physics-Informed Machine Learning Approach to Improve the Temperature Estimation Accuracy of Space-Resolved LPTNs. *2024 IEEE Transportation Electrification Conference and Expo (ITEC)*, 1–6. https://doi.org/10.1109/ITEC60657.2024.10599032

Sen, A., Ghajar-Rahimi, E., Aguirre, M., Navarro, L., Goergen, C. J., & Avril, S. (2024). Physics-Informed Graph Neural Networks to solve 1-D equations of blood flow. *Computer Methods and Programs in Biomedicine*, *257*, 108427. https://doi.org/10.1016/j.cmpb.2024.108427

Seshadri, A., & Lenin, N. C. (2020). Review based on losses, torque ripple, vibration and noise in switched reluctance motor. *IET Electric Power Applications*, *14*(8), 1311–1326. https://doi.org/10.1049/iet-epa.2019.0251

Seyyedi, A., Bohlouli, M., & Oskoee, S. N. (2023). Machine Learning and Physics: A Survey of Integrated Models. *ACM Comput. Surv.*, *56*(5), 115:1-115:33. https://doi.org/10.1145/3611383

Sharma, P., Chung, W. T., Akoush, B., & Ihme, M. (2023). A Review of Physics-Informed Machine Learning in Fluid Mechanics. *Energies*, *16*(5), Article 5. https://doi.org/10.3390/en16052343

Shokri, Gh., & Naderi, E. (2017). Research on simulation and modeling of simple and cost-effective BLDC motor drives. *International Journal of Modelling and Simulation*, *37*(1), 15–24. https://doi.org/10.1080/02286203.2016.1195665

Shukla, K., Xu, M., Trask, N., & Karniadakis, G. E. (2022). Scalable algorithms for physics-informed neural and graph networks. *Data-Centric Engineering*, *3*, e24. https://doi.org/10.1017/dce.2022.24

Soibam, J., Aslanidou, I., Kyprianidis, K., & Fdhila, R. B. (2024). Inverse flow prediction using ensemble PINNs and uncertainty quantification. *International Journal of Heat and Mass Transfer*, *226*, 125480. https://doi.org/10.1016/j.ijheatmasstransfer.2024.125480

Staessens, T., Lefebvre, T., & Crevecoeur, G. (2022). Adaptive Control of a Mechatronic System Using Constrained Residual Reinforcement Learning. *IEEE Transactions on Industrial Electronics*, *69*(10), 10447–10456. IEEE Transactions on Industrial Electronics. https://doi.org/10.1109/TIE.2022.3144565

Suhail, M., Akhtar, I., Kirmani, S., & Jameel, M. (2021). Development of Progressive Fuzzy Logic and ANFIS Control for Energy Management of Plug-In Hybrid Electric Vehicle. *IEEE Access*, *9*, 62219–62231. IEEE Access. https://doi.org/10.1109/ACCESS.2021.3073862

Suresh, S., & Rajeevan, P. P. (2020). Virtual Space Vector-Based Direct Torque Control Schemes for Induction Motor Drives. *IEEE Transactions on Industry Applications*, *56*(3), 2719–2728. IEEE Transactions on Industry Applications. https://doi.org/10.1109/TIA.2020.2978447

Tahkola, M., Keränen, J., Sedov, D., Far, M. F., & Kortelainen, J. (2020). Surrogate Modeling of Electrical Machine Torque Using Artificial Neural Networks. *IEEE Access*, *8*, 220027–220045. IEEE Access. https://doi.org/10.1109/ACCESS.2020.3042834

Taj, I., & Farooq, U. (2023). Towards Machine Learning-Based FPGA Backend Flow: Challenges and Opportunities. *Electronics*, *12*(4), Article 4. https://doi.org/10.3390/electronics12040935

Tan, C., Feng, Z.-M., Liu, X., Fan, J., Cui, W., Sun, R., & Ma, Q. (2020). Review of variable speed drive technology in beam pumping units for energy-saving. *Energy Reports*, *6*, 2676–2688. https://doi.org/10.1016/j.egyr.2020.09.018

Thangamuthu, A., Kumar, G., Bishnoi, S., Bhattoo, R., Krishnan, N. M. A., & Ranu, S. (2023). *Unravelling the Performance of Physics-informed Graph Neural Networks for Dynamical Systems* (No. arXiv:2211.05520). arXiv. https://doi.org/10.48550/arXiv.2211.05520

Tom, A. M., & Febin Daya, J. L. (2024). Machine learning techniques for vector control of permanent magnet synchronous motor drives. *Cogent Engineering*, *11*(1), 2323813. https://doi.org/10.1080/23311916.2024.2323813

Torchio, R., Conte, F., Martin, A., Bianchi, N., De Soricellis, M., Toso, F., Pase, F., Scarpa, M., Filippini, M., Lurtz, M., & Szepanski, D. (2025). Design and Experimental Validation of a Multiphysics Twin of a High-Voltage EV Motor. *IEEE Transactions on Transportation Electrification*, *11*(1), 3287–3297. IEEE Transactions on Transportation Electrification. https://doi.org/10.1109/TTE.2024.3437475

Torres-Tello, J., & Ko, S.-B. (2022). Optimizing a Multispectral-Images-Based DL Model, Through Feature Selection, Pruning and Quantization. *2022 IEEE International Symposium on Circuits and Systems (ISCAS)*, 1352–1356. https://doi.org/10.1109/ISCAS48785.2022.9937940

Traue, A., Book, G., Kirchgässner, W., & Wallscheid, O. (2022). Toward a Reinforcement Learning Environment Toolbox for Intelligent Electric Motor Control. *IEEE Transactions on Neural Networks and Learning Systems*, *33*(3), 919–928. IEEE Transactions on Neural Networks and Learning Systems. https://doi.org/10.1109/TNNLS.2020.3029573

Tripathi, S. M., & Vaish, R. (2019). Taxonomic research survey on vector controlled induction motor drives. *IET Power Electronics*, *12*(7), 1603–1615. https://doi.org/10.1049/iet-pel.2018.5216

Valencia, D. F., Tarvirdilu-Asl, R., Garcia, C., Rodriguez, J., & Emadi, A. (2021). Vision, Challenges, and Future Trends of Model Predictive Control in Switched Reluctance Motor Drives. *IEEE Access*, *9*, 69926–69937. IEEE Access. https://doi.org/10.1109/ACCESS.2021.3078366

Wallscheid, O. (2021). Thermal Monitoring of Electric Motors: State-of-the-Art Review and Future Challenges. *IEEE Open Journal of Industry Applications*, *2*, 204–223. IEEE Open Journal of Industry Applications. https://doi.org/10.1109/OJIA.2021.3091870

Wang, F., Mei, X., Rodriguez, J., & Kennel, R. (2017). Model predictive control for electrical drive systems-an overview. *CES Transactions on Electrical Machines and Systems*, *1*(3), 219–230. CES Transactions on Electrical Machines and Systems. https://doi.org/10.23919/TEMS.2017.8086100

Wang, S., Teng, Y., & Perdikaris, P. (2021). Understanding and Mitigating Gradient Flow Pathologies in Physics-Informed Neural Networks. *SIAM Journal on Scientific Computing*, *43*(5), A3055–A3081. https://doi.org/10.1137/20M1318043

Wang, Y., Tian, Y., Kirk, T., Laris, O., Ross, J. H., Noebe, R. D., Keylin, V., & Arróyave, R. (2020). Accelerated design of Fe-based soft magnetic materials using machine learning and stochastic optimization. *Acta Materialia*, *194*, 144–155. https://doi.org/10.1016/j.actamat.2020.05.006

Warren, P., Ali, H., Ebrahimi, H., & Ghosh, R. (2021, September 16). *Rapid Defect Detection and Classification in Images Using Convolutional Neural Networks*. ASME Turbo Expo 2021: Turbomachinery Technical Conference and Exposition. https://doi.org/10.1115/GT2021-59801

Watthewaduge, G., Sayed, E., Emadi, A., & Bilgin, B. (2020). Electromagnetic Modeling Techniques for Switched Reluctance Machines: State-of-the-Art Review. *IEEE Open Journal of the Industrial Electronics Society*, *1*, 218–234. IEEE Open Journal of the Industrial Electronics Society. https://doi.org/10.1109/OJIES.2020.3016242

Wu, Y., Sicard, B., & Gadsden, S. A. (2024). Physics-informed machine learning: A comprehensive review on applications in anomaly detection and condition monitoring. *Expert Systems with Applications*, *255*, 124678. https://doi.org/10.1016/j.eswa.2024.124678

Wu, Z., Wang, H., He, C., Zhang, B., Xu, T., & Chen, Q. (2023). The Application of Physics-Informed Machine Learning in Multiphysics Modeling in Chemical Engineering. *Industrial & Engineering Chemistry Research*, *62*(44), 18178–18204. https://doi.org/10.1021/acs.iecr.3c02383

Xie, C., Du, S., Wang, J., Lao, J., & Song, H. (2023). Intelligent modeling with physics-informed machine learning for petroleum engineering problems. *Advances in Geo-Energy Research*, *8*(2), Article 2. https://doi.org/10.46690/ager.2023.05.01

Xu, Y., Kohtz, S., Boakye, J., Gardoni, P., & Wang, P. (2023). Physics-informed machine learning for reliability and systems safety applications: State of the art and challenges. *Reliability Engineering & System Safety*, *230*, 108900. https://doi.org/10.1016/j.ress.2022.108900

Xu, Z., Li, T., Zhang, F., Zhang, Y., Lee, D.-H., & Ahn, J.-W. (2022). A Review on Segmented Switched Reluctance Motors. *Energies*, *15*(23), Article 23. https://doi.org/10.3390/en15239212

Xue, Z., Niu, S., Chau, A. M. H., Luo, Y., Lin, H., & Li, X. (2023). Recent Advances in Multi-Phase Electric Drives Model Predictive Control in Renewable Energy Application: A State-of-the-Art Review. *World Electric Vehicle Journal*, *14*(2), Article 2. https://doi.org/10.3390/wevj14020044

Yang, C., Chen, C., He, W., Cui, R., & Li, Z. (2019). Robot Learning System Based on Adaptive Neural Control and Dynamic Movement Primitives. *IEEE Transactions on Neural Networks and Learning Systems*, *30*(3), 777–787. IEEE Transactions on Neural Networks and Learning Systems. https://doi.org/10.1109/TNNLS.2018.2852711

Yazdani, S., & Tahani, M. (2024). Data-driven discovery of turbulent flow equations using physics-informed neural networks. *Physics of Fluids*, *36*(3), 035107. https://doi.org/10.1063/5.0190138

Yin, J., Li, Y., & Yue, S. (2024). Comprehensive Investigation on Iteration Algorithms for Solving Nonlinear Magnetic Field Problems. *IEEE Transactions on Magnetics*, *60*(9), 1–5. IEEE Transactions on Magnetics. https://doi.org/10.1109/TMAG.2024.3404601

Yousef, H. A., Hamdy, M., Saleem, A., Nashed, K., Mesbah, M., & Shafiq, M. (2019). Enhanced adaptive control for a benchmark piezoelectric-actuated system via fuzzy approximation. *International Journal of Adaptive Control and Signal Processing*, *33*(9), 1329–1343. https://doi.org/10.1002/acs.3033

Zargari, N., Cheng, Z., & Paes, R. (2017). A guide to matching medium voltage drive topology to petrochemical applications. *2017 Petroleum and Chemical Industry Technical Conference (PCIC)*, 133–142. https://doi.org/10.1109/PCICON.2017.8188732

Zhang, H., Jiang, L., Chu, X., Wen, Y., Li, L., Liu, J., Xiao, Y., & Wang, L. (2025). Combining physics-informed graph neural network and finite difference for solving forward and inverse spatiotemporal PDEs. *Computer Physics Communications*, *308*, 109462. https://doi.org/10.1016/j.cpc.2024.109462

Zhang, S., Wallscheid, O., & Porrmann, M. (2023). Machine Learning for the Control and Monitoring of Electric Machine Drives: Advances and Trends. *IEEE Open Journal of Industry Applications*, *4*, 188–214. IEEE Open Journal of Industry Applications. https://doi.org/10.1109/OJIA.2023.3284717

Zhao, S., Peng, Y., Zhang, Y., & Wang, H. (2022). Parameter Estimation of Power Electronic Converters With Physics-Informed Machine Learning. *IEEE Transactions on Power Electronics*, *37*(10), 11567–11578. IEEE Transactions on Power Electronics. https://doi.org/10.1109/TPEL.2022.3176468

Zhao, T. (2023). FPGA-Based Machine Learning: Platforms, Applications, Design Considerations, Challenges, and Future Directions. *Highlights in Science, Engineering and Technology*, *62*, 96–101. https://doi.org/10.54097/hset.v62i.10430

Zhao, X., Shirvan, K., Salko, R. K., & Guo, F. (2020). On the prediction of critical heat flux using a physics-informed machine learning-aided framework. *Applied Thermal Engineering*, *164*, 114540. https://doi.org/10.1016/j.applthermaleng.2019.114540

Zhao, Y., Chen, W., & Yang, X. (2024). Adaptive Sampling Stochastic Multigradient Algorithm for Stochastic Multiobjective Optimization. *Journal of Optimization Theory and Applications*, *200*(1), 215–241. https://doi.org/10.1007/s10957-023-02334-w

Zhu, Z. Q., Liang, D., & Liu, K. (2021). Online Parameter Estimation for Permanent Magnet Synchronous Machines: An Overview. *IEEE Access*, *9*, 59059–59084. IEEE Access. https://doi.org/10.1109/ACCESS.2021.3072959

Zideh, M. J., Chatterjee, P., & Srivastava, A. K. (2024). Physics-Informed Machine Learning for Data Anomaly Detection, Classification, Localization, and Mitigation: A Review, Challenges, and Path Forward. *IEEE Access*, *12*, 4597–4617. IEEE Access. https://doi.org/10.1109/ACCESS.2023.3347989

Zonta, T., da Costa, C. A., da Rosa Righi, R., de Lima, M. J., da Trindade, E. S., & Li, G. P. (2020). Predictive maintenance in the Industry 4.0: A systematic literature review. *Computers & Industrial Engineering*, *150*, 106889. https://doi.org/10.1016/j.cie.2020.106889